\documentclass{llncs}
\usepackage{amsmath,amssymb,amsfonts}
\usepackage{multirow,multicol}
\usepackage{tabulary,colortbl}
\usepackage[mathscr]{eucal}
\usepackage[ruled]{algorithm}
\usepackage{algpseudocode}
\usepackage{booktabs}
\usepackage{url}
\usepackage[final]{graphicx}
\usepackage[braket]{qcircuit}
\usepackage{caption}
\usepackage{booktabs}
\usepackage{listings}
\usepackage{upquote}
\usepackage{color}
\usepackage{a4wide}

\definecolor{greencomments}{rgb}{0,0.5,0}

\newcommand{\Z}{\ensuremath{\mathbb{Z}}}
\newcommand{\F}{\ensuremath{\mathbb{F}}}

\mathchardef\mhyphen="2D

\renewcommand{\O}{\mathcal{O}}

\newcommand{\dbl}{\mathrm{dbl}}

\newcommand{\inv}{\mathtt{inv}}
\newcommand{\mul}{\mathtt{mul}}
\newcommand{\squ}{\mathtt{squ}}
\newcommand{\nega}{\mathtt{neg}}

\newcommand{\liq}{LIQ$Ui|\rangle$}

\lstdefinelanguage{FSharp}%
{morekeywords={let, new, match, with, rec, open, module, namespace, type, of, member, %
and, for, while, true, false, in, do, begin, fun, function, return, yield, try, %
mutable, if, then, else, cloud, async, static, use, abstract, interface, inherit, finally },
  otherkeywords={ let!, return!, do!, yield!, use!, var, from, select, where, order, by },
  keywordstyle=\color{blue},
  sensitive=true,
  basicstyle=\ttfamily,
	breaklines=true,
  xleftmargin=\parindent,
  aboveskip=\bigskipamount,
	tabsize=4,
  morecomment=[l][\color{greencomments}]{///},
  morecomment=[l][\color{greencomments}]{//},
  morecomment=[s][\color{greencomments}]{{(*}{*)}},
  morestring=[b]",
  showstringspaces=false,
  literate={`}{\`}1,
  stringstyle=\color{red},
}

\newcommand{\ctr}{\mathrm{ctrl}}
\newcommand{\tmp}{\mathrm{tmp}}
\makeatletter
\def\mop#1{\mathop{\operator@font {#1\null}}}
\makeatother

\def\ord{{\mop{ord}}}

\newcommand{\triang}{\quad\,\blacktriangleright\ \,}

\newcommand{\gateout}[1]{*+<.6em>{#1\ } \POS ="i",[0,0]+R *{\triang},"i"+UR;"i"+UL **\dir{-};"i"+DL **\dir{-};"i"+DR **\dir{-};"i"+UR **\dir{-},"i" \qw}
\newcommand{\multigateout}[2]{*+<1em,.9em>{\hphantom{#2}} \POS [0,0]="i",[0,0].[#1,0]="e", !C *{#2}, [0,0]+R *{\triang}, "e"+UR;"e"+UL **\dir{-};"e"+DL **\dir{-};"e"+DR **\dir{-};"e"+UR **\dir{-},"i" \qw}
\newcommand{\multigateoutlow}[2]{*+<1em,.9em>{\hphantom{#2}} \POS [0,0]="i",[0,0].[#1,0]="e", !C *{#2}, [#1,0]+R *{\triang}, "e"+UR;"e"+UL **\dir{-};"e"+DL **\dir{-};"e"+DR **\dir{-};"e"+UR **\dir{-},"i" \qw}
\newcommand{\multigateoutsecond}[2]{*+<1em,.9em>{\hphantom{#2}} \POS [0,0]="i",[0,0].[#1,0]="e", !C *{#2}, [1,0]+R *{\triang}, "e"+UR;"e"+UL **\dir{-};"e"+DL **\dir{-};"e"+DR **\dir{-};"e"+UR **\dir{-},"i" \qw}
\newcommand{\multigateouttwo}[2]{*+<1em,.9em>{\hphantom{#2}} \POS [0,0]="i",[0,0].[#1,0]="e", !C *{#2}, [0,0]+R *{\triang}, [#1,0]+R *{\triang}, "e"+UR;"e"+UL **\dir{-};"e"+DL **\dir{-};"e"+DR **\dir{-};"e"+UR **\dir{-},"i" \qw}
\newcommand{\multigateouttwolow}[2]{*+<1em,.9em>{\hphantom{#2}} \POS [0,0]="i",[0,0].[#1,0]="e", !C *{#2}, [1,0]+R *{\triang},  [#1,0]+R *{\triang},"e"+UR;"e"+UL **\dir{-};"e"+DL **\dir{-};"e"+DR **\dir{-};"e"+UR **\dir{-},"i" \qw}
\newcommand{\multigateouttwohi}[2]{*+<1em,.9em>{\hphantom{#2}} \POS [0,0]="i",[0,0].[#1,0]="e", !C *{#2}, [0,0]+R *{\triang},  [1,0]+R *{\triang},"e"+UR;"e"+UL **\dir{-};"e"+DL **\dir{-};"e"+DR **\dir{-};"e"+UR **\dir{-},"i" \qw}
\newcommand{\multigateoutthreelow}[2]{*+<1em,.9em>{\hphantom{#2}} \POS [0,0]="i",[0,0].[#1,0]="e", !C *{#2}, [1,0]+R *{\triang}, [2,0]+R *{\triang}, [3,0]+R *{\triang}, "e"+UR;"e"+UL **\dir{-};"e"+DL **\dir{-};"e"+DR **\dir{-};"e"+UR **\dir{-},"i" \qw}
\newcommand{\multigateoutthreeup}[2]{*+<1em,.9em>{\hphantom{#2}} \POS [0,0]="i",[0,0].[#1,0]="e", !C *{#2}, [0,0]+R *{\triang}, [1,0]+R *{\triang}, [2,0]+R *{\triang}, "e"+UR;"e"+UL **\dir{-};"e"+DL **\dir{-};"e"+DR **\dir{-};"e"+UR **\dir{-},"i" \qw}

\title{
Quantum Resource Estimates for Computing \\Elliptic Curve Discrete Logarithms 
}

\author{Martin Roetteler \and Michael Naehrig \and Krysta M.~Svore \and Kristin Lauter}
\institute{Microsoft Research, USA}

\begin{document}
\maketitle

\begin{abstract}
We give precise quantum resource estimates for Shor's algorithm to compute discrete logarithms on elliptic curves over prime fields. The estimates are derived from a simulation of a Toffoli gate network for controlled elliptic curve point addition, implemented within the framework of the quantum computing software tool suite \liq. We determine circuit implementations for reversible modular arithmetic, including modular addition, multiplication and inversion, as well as reversible elliptic curve point addition. We conclude that elliptic curve discrete logarithms on an elliptic curve defined over an $n$-bit  prime field can be computed on a quantum computer with at most $9n + 2\lceil\log_2(n)\rceil+10$ qubits using a quantum circuit of at most $448 n^3 \log_2(n) + 4090 n^3$ Toffoli gates. We are able to classically simulate the Toffoli networks corresponding to the controlled elliptic curve point addition as the core piece of Shor's algorithm for the NIST standard curves P-192, P-224, P-256, P-384 and P-521. Our approach allows gate-level comparisons to recent resource estimates for Shor's factoring algorithm. The results also support estimates given earlier by Proos and Zalka and indicate that, for current parameters at comparable classical security levels, the number of qubits required to tackle elliptic curves is less than for attacking RSA, suggesting that indeed ECC is an easier target than RSA. 

\end{abstract}

%%%%%%%%%%%%%%%%%%%%%%%%%%%%%%%%%%%%%%%%%%%%%%%%%%%%%%%%%%%%%%%%%%%%%%%%%%%%%%%
\section{ Introduction}
\label{sec:intro}

\subsubsection{Elliptic curve cryptography (ECC).}
Elliptic curves are a fundamental building block of today's cryptographic landscape. Thirty years after their introduction to cryptography~\cite{miller85,koblitz87}, they are used to instantiate public key mechanisms such as key exchange~\cite{DH76} and digital signatures~\cite{ElGamal84,ECDSA} that are widely deployed in various cryptographic systems. Elliptic curves are used in applications such as transport layer security~\cite{rfc5246,rfc4492}, secure shell~\cite{rfc5656}, the Bitcoin digital currency system~\cite{bitcoin}, in national ID cards~\cite{burgerkarte}, the Tor anonymity network~\cite{Tor}, and the WhatsApp messaging app~\cite{whatsapp}, just to name a few. Hence, they play a significant role in securing our data and communications. 

Different standards (e.g., \cite{SEC2,fips186-4}) and standardization efforts (e.g., \cite{brainpool,cfrg}) have identified elliptic curves of different sizes targeting different levels of security. Notable curves with widespread use are the NIST curves P-256, P-384, P-521, which are curves in Weierstrass form over special primes of size 256, 384, and 521 bits respectively, the Bitcoin curve \texttt{secp256k1} from the SEC2~\cite{SEC2} standard and the Brainpool curves~\cite{brainpool}. More recently, Bernstein's Curve25519~\cite{curve25519}, a Montgomery curve over a 255-bit prime field, has seen more and more deployment, and it has been recommended to be used in the next version of the TLS protocol~\cite{rfc7748} along with another even more recent curve proposed by Hamburg called Goldilocks~\cite{goldilocks}.

The security of elliptic curve cryptography relies on the hardness of computing discrete logarithms in elliptic curve groups, i.e. the difficulty of the Elliptic Curve Discrete Logarithm Problem (ECDLP). Elliptic curves have the advantage of relatively small parameter and key sizes in comparison to other cryptographic schemes, such as those based on RSA~\cite{RSA78} or finite field discrete logarithms~\cite{DH76}, when compared at the same security level. For example, according to NIST recommendations from 2016, a 256-bit elliptic curve provides a similar resistance against classical attackers as an RSA modulus of size 3072 bits\footnote{Opinions about such statements of equivalent security levels differ, for an overview see~\url{https://www.keylength.com}. There is consensus about the fact that elliptic curve parameters can be an order of magnitude smaller than parameters for RSA or finite field discrete logarithm systems to provide similar security.}. This advantage arises from the fact that the currently known best algorithms to compute elliptic curve discrete logarithms are exponential in the size of the input parameters\footnote{For a recent survey, see~\cite{GG16}.}, whereas there exist subexponential algorithms for factoring~\cite{LL93,CP05} and finite field discrete logarithms~\cite{Gordon93,JL03}. 

\subsubsection{The quantum computer threat.}
In his famous paper~\cite{shor}, Peter Shor presented two polynomial-time quantum algorithms, one for integer factorization and another one for computing discrete logarithms in a finite field of prime order. Shor notes that the latter algorithm can be generalized to other fields. It also generalizes to the case of elliptic curves. Hence, given the prerequisite that a large enough general purpose quantum computer can be built, the algorithms in Shor's paper completely break all current crypto systems based on the difficulty of factoring or computing discrete logarithms. Scaling up the parameters for such schemes to sizes for which Shor's algorithm becomes practically infeasible will most likely lead to highly impractical instantiations.

Recent years have witnessed significant advances in the state of quantum computing hardware. Companies have invested in the development of qubits, and the field has seen an emergence of startups, with some focusing on quantum hardware, others on software for controlling quantum computers, and still others offering consulting services to ready for the quantum future.
The predominant approach to quantum computer hardware focuses on physical implementations that are scalable, digital, programmable, and universal. With the amount of investment in quantum computing hardware, the pace of scaling is increasing and underscoring the need to understand the scaling of the difficulty of ECDLP.

\subsubsection{Language-Integrated Quantum Operations: \liq.}
As quantum hardware advances towards larger-scale systems of upwards of tens to hundreds of qubits, there is a critical need for a software architecture to program and control the device.  We use the \liq\ software architecture \cite{liquid} to determine the resource costs of solving the ECDLP. \liq\ is a high-level programming language for quantum algorithms embedded in F\#, a compilation stack to translate and compile quantum algorithms into quantum circuits, and a simulator to test and run quantum circuits\footnote{See \url{http://stationq.github.io/Liquid/} and \url{https://github.com/StationQ/Liquid}.}. \liq\ can simulate roughly 32 qubits in 32GB RAM, however, we make use of the fact that reversible circuits can be simulated efficiently on classical input states for thousands of qubits. 

\subsubsection{Gate sets and Toffoli gate networks.} The basic underlying fault-tolerant architecture and coding scheme of a quantum computer determine the universal gate set, and hence by extension also the synthesis problems that have to be solved in order to compile high-level, large-scale algorithms into a sequence of operations that an actual physical quantum computer can then execute. A gate set that arises frequently and that has been studied often in the literature, but by no means the only conceivable gate set, is the so-called Clifford$+T$ gate set \cite{NC:2000}.
This gate set consists of the Hadamard gate $H=\text{\footnotesize $\frac{1}{\sqrt{2}}\left[\begin{smallmatrix} 1 & 1\\ 1 & -1 \end{smallmatrix}\right]$}$, the phase gate $P={\rm diag}(1,i)$, and the controlled NOT (CNOT) gate which maps $(x,y) \mapsto (x, x\oplus y)$ as generators of the Clifford group, along with the $T$ gate given by $T={\rm diag}(1,\exp(\pi i/4))$. The Clifford$+T$ gate set is known to be universal \cite{NC:2000}. This means that it can be used to approximate any given target unitary single qubit operation to within precision $\varepsilon$ using sequences of length $4 \log_2({1/\varepsilon})$ \cite{Selinger,KMM:2016}, and using an entangling gate such as the CNOT gate, the Clifford$+T$ gate set can approximate any unitary operation. 
When assessing the complexity of a quantum circuit built from Clifford$+T$ gates, often only $T$-gates are counted as many fault-tolerant implementations of the Clifford$+T$ gate set at the logical gate level require much more resources for $T$-gates than for Clifford gates \cite{Fowl12f}. 

In this paper, we base reversible computations entirely on the Toffoli gate. The Toffoli gate $\ket{x,y,z} \mapsto \ket{x,y,z\oplus xy}$ is known to be universal for reversible computing \cite{NC:2000} and can  be implemented exactly over the Clifford$+T$ gate set, see \cite{Selinger:2013} for a $T$-depth $1$ implementation using a total of $7$ qubits and \cite{Amy:2013} for a $T$-depth $3$ realization using a total of $3$ qubits. As discussed in~\cite[Section~V]{HRS16}, there are two main reasons for focusing on Toffoli gate networks as our preferred realization of quantum circuits. The first is that because the Toffoli gate can be implemented exactly over the Clifford$+T$ gate set, Toffoli networks do not have gate synthesis overhead. The second is testability and debugging. Toffoli gate networks can be simulated using classical reversible simulators. While a fully functional simulation of a quantum circuit could be deemed feasible for circuits on up to 50 qubits, classical simulation of Toffoli gate-based circuits can deal with a lot more qubits. Also, for implementations on actual quantum hardware, Toffoli gate circuits can be debugged efficiently and faults can be localized through binary search \cite{HRS16}.

\subsubsection{Estimating quantum resources for Shor's ECDLP algorithm.}
Understanding the concrete requirements for a quantum computer that is able to run Shor's algorithm helps to put experimental progress in quantum computing into perspective. Although it is clear that the polynomial runtime asymptotically breaks ECC, constant factors can make an important difference when actually implementing the algorithm. 

In~\cite{PZ03}, Proos and Zalka describe how Shor's algorithm can be implemented for the case of elliptic curve groups. They conclude with a table of resource estimates for the number of logical qubits and time (measured in ``1-qubit additions'') depending on the bitsize of the elliptic curve. Furthermore, they compare these estimates to those for Shor's factoring algorithm and argue that computing elliptic curve discrete logarithms is significantly easier than factoring RSA moduli at comparable classical security levels.
However, some questions remained unanswered by \cite{PZ03}, the most poignant of which being whether it is actually possible to construct and simulate the circuits to perform elliptic curve point addition in order to get confidence in their correctness. Another question that remained open is whether it is possible to determine constants that were left in terms of asymptotic scaling and whether some of the proposed circuit constructions to compress registers and to synchronize computations can actually be implemented in code that can then be automatically generated for arbitrary input curves. 

Here we build on their work and fully program and simulate the underlying arithmetic.  We verify the correctness of our algorithms and obtain concrete resource costs measured by the overall number of logical qubits, the number of Toffoli gates and the depth of a quantum circuit for implementing Shor's algorithm.

\subsubsection{Contributions.}
In this paper, we present precise resource estimates for quantum circuits that implement Shor's algorithm to solve the ECDLP. In particular, our contributions are as follows:
\begin{itemize}
\item We describe reversible algorithms for modular quantum arithmetic. This includes modular addition, subtraction, negation and doubling of integers held in quantum registers, modular multiplication, squaring and inversion. 
\item For modular multiplication, we consider two different approaches, besides an algorithm based on modular doublings and modular additions, we also give a circuit for Montgomery multiplication.
\item Based on our implementations it transpired that using Montgomery arithmetic is beneficial as the cost for the multiplication can be seen to be lower than that of the double-and-add method. The latter requires less ancillas, however, in the given algorithm there are always enough ancillas available as overall a relatively large number of ancillas must be provided. 
\item Our modular inversion algorithm is a reversible implementation of the Montgomery inverse via the binary extended Euclidean (binary GCD) algorithm. To realize this algorithm as a circuit, we introduce tools that might be of independent interest for other reversible algorithms.
\item We describe a quantum circuit for elliptic curve point addition in affine coordinates and describe how it can be used to implement scalar multiplication within Shor's algorithm.
\item We have implemented all of the above algorithms in F\# within the framework of the quantum computing software tool suite \liq\ \cite{liquid} and have simulated and tested all of these algorithms for real-world parameters of up to $521$ bits\footnote{Our code will be made publicly available at \url{http://microsoft.com/quantum}.}.
\item  Derived from our implementation, we present concrete resource estimates for the total number of qubits, the number of Toffoli gates and the depth of the Toffoli gate networks to realize Shor's algorithm and its subroutines.
We compare the quantum resources for solving the ECDLP to those required in Shor's factoring algorithm that were obtained in the recent work~\cite{HRS16}.
\end{itemize}

\subsubsection{Results.}
Our implementation realizes a reversible circuit for controlled elliptic curve point addition on an elliptic curve defined over a field of prime order with $n$ bits and needs at most $9n + 2\lceil\log_2(n)\rceil+10$ qubits. An interpolation of the data points for the number of Toffoli gates shows that the quantum circuit can be implemented with at most roughly $224 n^2 \log_2(n) + 2045 n^2$ Toffoli gates. For Shor's full algorithm, the point addition needs to be run $2n$ times sequentially and does not need additional qubits. The overall number of Toffoli gates is thus about $448 n^3 \log_2(n) + 4090n^3$. For example, our simulation of the point addition quantum circuit for the NIST standardized curve P-256 needs 2330 logical qubits and the full Shor algorithm would need about $1.26\cdot 10^{11}$ Toffoli gates. In comparison, Shor's factoring algorithm for a 3072-bit modulus needs 6146 qubits and $1.86\cdot 10^{13}$ Toffoli gates\footnote{These estimates are interpolated from the results in~\cite{HRS16}.}, which aligns with results by Proos and Zalka showing that it is easier to break ECC than RSA at comparable classical security.

Our estimates provide a data point that allows a better understanding of the requirements to run Shor's quantum ECDLP algorithm and we hope that they will serve as a basis to make better predictions about the time horizon until which elliptic curve cryptography can still be considered secure. 
Besides helping to gain a better understanding of the post-quantum (in-) security of elliptic curve cryptosystems, we hope that our reversible algorithms (and their \liq\ implementations) for modular arithmetic and the elliptic curve group law are of independent interest to some, and might serve as building blocks for other quantum algorithms.

\section{Elliptic curves and Shor's algorithm}
\label{sec:ecshor}
This section provides some background on elliptic curves over finite fields, the elliptic curve discrete logarithm problem (ECDLP) and Shor's quantum algorithm to solve the ECDLP. Throughout, we restrict to the case of curves defined over prime fields of large characteristic. 

\subsection{Elliptic curves and the ECDLP}
Let $p>3$ be a prime. Denote by $\F_p$ the finite field with $p$ elements. An elliptic curve over $\F_p$ is a projective, non-singular curve of genus 1 with a specified base point. It can be given by an affine Weierstrass model, i.e. it can be viewed as the set of all solutions $(x,y)$ to the equation $E: y^2 = x^3 + ax + b$ with two curve constants $a,b\in \F_p$, together with a point at infinity $\O$. The set of $\F_p$-rational points consists of $\O$ and all solutions $(x,y)\in \F_p\times \F_p$ and is denoted by $E(\F_p) = \{(x,y) \in \F_p\times \F_p \mid y^2 = x^3 + ax + b\}\cup \{\O\}$. The set $E(\F_p)$ is an abelian group with respect to a group operation ``$+$'' that is defined via rational functions in the point coordinates with $\O$ as the neutral element. Similarly, for a field extension $\F\supseteq \F_p$, one similarly defines the group of $\F$-rational points $E(\F)$ and if $\F$ is an algebraic closure of $\F_p$, we simply denote $E = E(\F)$. For an extensive treatment of elliptic curves, we refer the reader to~\cite{silverman}.

The elliptic curve group law on an affine Weierstrass curve can be computed as follows. Let $P_1,P_2\in E$ and let $P_3 = P_1 + P_2$. If $P_1 = \O$ then $P_3 = P_2$ and if $P_2 = \O$, then $P_3 = P_1$. Now let $P_1 \neq \O \neq P_2$ and write $P_1 = (x_1, y_1)$ and $P_2 = (x_2, y_2)$ for $x_1,y_1,x_2,y_2 \in \F$. If $P_2 = -P_1$, then $x_1 = x_2$, $y_2 = -y_1$ and $P_3 = \O$. If neither of the previous cases occurs, then $P_3 = (x_3, y_3)$ is an affine point and can be computed as
\[
x_3 = \lambda^2 - x_1 - x_2,\quad y_3 = (x_1 - x_3) \lambda - y_1,
\] 
where $\lambda = \frac{y_2 - y_1}{x_2 - x_1}$ if $P_1 \neq P_2$, i.e. $x_1 \neq x_2$, and $\lambda = \frac{3x_1^2 + a}{2y_1}$ if $P_1 = P_2$. For a positive integer $m$, denote by $[m]P$ the $m$-fold sum of $P$, i.e. $[m]P = P + \cdots + P$, where $P$ occurs $m$ times. Extended to all $m\in\Z$ by $[0]P = \O$ and $[-m]P = [m](-P)$, the map $[m]: E \rightarrow E, P \mapsto [m]P$ is called the multiplication-by-$m$ map or simply scalar multiplication by $m$. Scalar multiplication (or group exponentiation in the multiplicative setting) is one of the main ingredients for discrete-logarithm-based cryptographic protocols. It is also an essential operation in Shor's ECDLP algorithm. The order $\mathrm{ord}(P)$ of a point $P$ is the smallest positive integer $r$ such that $[r]P=\O$.

Curves that are most widely used in cryptography are defined over large prime fields. One works in a cyclic subgroup of $E(\F_p)$ of large prime order $r$, where $\#E(\F_p) = h\cdot r$. The group order can be written as $\#E(\F_p) = p + 1 - t$, where $t$ is called the trace of Frobenius and the Hasse bound ensures that $|t| \leq 2\sqrt{p}$. Thus $\#E(\F_p)$ and $p$ are of roughly the same size. The most efficient instantiations of ECC are achieved for small cofactors $h$. For example, the above mentioned NIST curves have prime order, i.e. $h=1$, and Curve25519 has cofactor $h=8$.
Let $P\in E(\F_p)$ be an $\F_p$-rational point on $E$ of order $r$ and let $Q \in \langle P \rangle$ be an element of the cyclic subgroup generated by $P$. The Elliptic Curve Discrete Logarithm Problem (ECDLP) is the problem to find the integer $m \in \Z/r\Z$ such that 
$Q=[m]P$. 
The bit security of an elliptic curve is estimated by extrapolating the runtime of the most efficient algorithms for the ECDLP.

The currently best known classical algorithms to solve the ECDLP are based on parallelized versions of Pollard's rho algorithm~\cite{Pollard78,OW99,Pollard00}. When working in a group of order $n$, the expected running time for solving a single ECDLP is $(\sqrt{\pi/2}+o(1))\sqrt{n}$ group operations based on the birthday paradox. This is exponential in the input size $\log(n)$. See~\cite{GG16} for further details and~\cite{BCM14} for a concrete, implementation-based security assessment. 

\subsection{Shor's quantum algorithm for solving the ECDLP}
In~\cite{shor}, Shor presented two polynomial time quantum algorithms, one for factoring integers, the other for computing discrete logarithms in finite fields. The second one can naturally be applied for computing discrete logarithms in the group of points on an elliptic curve defined over a finite field.

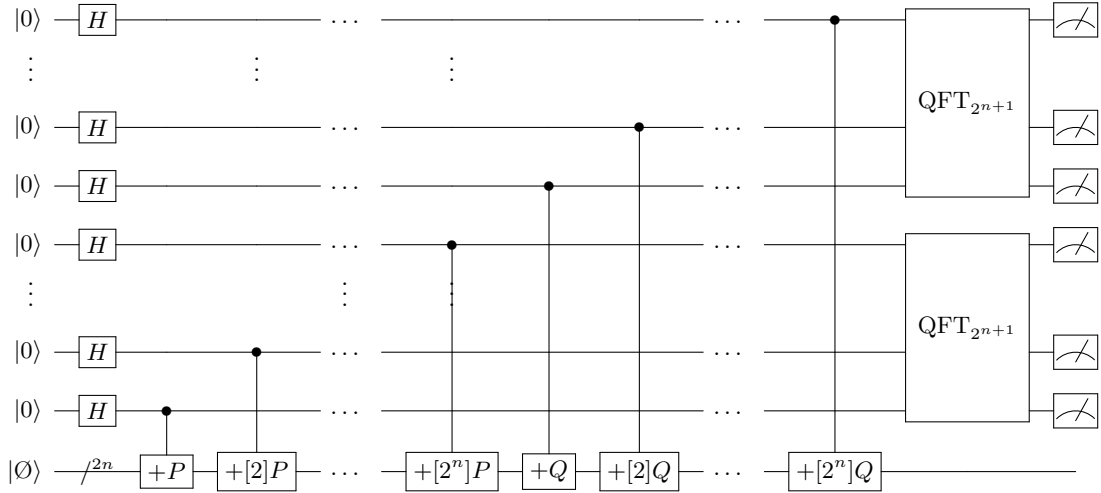
\begin{figure}[hbt]
{\footnotesize
\Qcircuit @C=1.0em @R=1.0em { 
\push{\rule{1.2em}{0em}}& \lstick{\ket{0}} & \gate{H} & \qw & \qw & \qw & \dots  & \push{\rule{0.5em}{0em}} & \qw & \qw & \qw & \qw & \dots & \push{\rule{0.5em}{0em}} & \ctrl{10} &  \multigate{4}{\mathrm{QFT}_{2^{n+1}}} & \meter\\
& \lstick{\vdots\ } & & & \vdots & & & & \vdots & & & & & & & & \\
& & & & & & & & & & & & & & & & \\
& \lstick{\ket{0}} & \gate{H} & \qw & \qw & \qw & \dots & \push{\rule{0.5em}{0em}} & \qw & \qw & \ctrl{7} & \qw & \dots & \push{\rule{0.5em}{0em}} & \qw  & \ghost{\mathrm{QFT}_{2^{n+1}}} & \meter\\
& \lstick{\ket{0}} & \gate{H} & \qw & \qw & \qw & \dots & \push{\rule{0.5em}{0em}} & \qw & \ctrl{6} & \qw & \qw & \dots & \push{\rule{0.5em}{0em}} & \qw  & \ghost{\mathrm{QFT}_{2^{n+1}}} & \meter\\
& \lstick{\ket{0}} & \gate{H} & \qw & \qw & \qw & \dots & \push{\rule{0.5em}{0em}} & \ctrl{5} & \qw & \qw & \qw & \dots & \push{\rule{0.5em}{0em}} & \qw  &  \multigate{4}{\mathrm{QFT}_{2^{n+1}}} & \meter\\
& \lstick{\vdots\ } & & & & & \vdots & & \vdots & & & & & & & &\\
& & & & & & & & & & & & & & & &\\
& \lstick{\ket{0}} & \gate{H} & \qw & \ctrl{2} & \qw & \dots & \push{\rule{0.5em}{0em}} & \qw & \qw & \qw & \qw & \dots & \push{\rule{0.5em}{0em}} & \qw  & \ghost{\mathrm{QFT}_{2^{n+1}}} & \meter\\
& \lstick{\ket{0}} & \gate{H} & \ctrl{1} & \qw & \qw & \dots  & \push{\rule{0.5em}{0em}} & \qw & \qw & \qw & \qw & \dots & \push{\rule{0.5em}{0em}} & \qw  & \ghost{\mathrm{QFT}_{2^{n+1}}} & \meter\\
& \lstick{\ket{\O}} & \qw \slash^{2n} & \gate{+P} & \gate{+[2]P} & \qw  & \dots & \push{\rule{0.5em}{0em}} & \gate{+[2^n]P} & \gate{+Q} & \gate{+[2]Q} & \qw & \dots & \push{\rule{0.5em}{0em}} & \gate{+[2^n]Q} & \qw & \qw
}
}
\caption{\label{fig:shorcircuit} Shor's algorithm to compute the discrete logarithm in the subgroup of an elliptic curve generated by a point $P$. The input to the problem is a point $Q$, and the task is to find $m\in\{1,\dots,\ord(P)\}$ such that $Q = [m]P$. The circuit naturally decomposes into three parts, namely (i) the Hadamard layer on the left, (ii) a double scalar multiplication (in this figure implemented as a cascade of conditional point additions), and (iii) the quantum Fourier transform ${\rm QFT}$ and subsequent measurement in the standard basis which is performed at the end.
}
\end{figure}

We are given an instance of the ECDLP as described above. Let $P\in E(\F_p)$ be a fixed generator of a cyclic subgroup of $E(\F_p)$ of known order $\ord(P) = r$, let  $Q \in \langle P \rangle$ be a fixed element in the subgroup generated by $P$; our goal is to find the unique integer $m\in\{1,\dots, r\}$ such that $Q=[m]P$. 
Shor's algorithm proceeds as follows. First, two registers of length $n+1$ qubits\footnote{Hasse's bound guarantees that the order of $P$ can be represented with $n+1$ bits.} are created and each qubit is initialized in the $\ket{0}$ state. Then a Hadamard transform $H$ is applied to each qubit, resulting in the state $\frac{1}{2^{n+1}} \sum_{k,\ell=0}^{2^{n+1}-1} \ket{k,\ell}$. Next, conditioned on the content of the register holding the label $k$ or $\ell$, we add the corresponding multiple of $P$ and $Q$, respectively, i.\,e., we 
implement the map 
\[ 
\frac{1}{2^{n+1}} \sum_{k,\ell=0}^{2^{n+1}-1} \ket{k,\ell} \mapsto 
\frac{1}{2^{n+1}} \sum_{k,\ell=0}^{2^{n+1}-1} \ket{k,\ell}\ket{[k] P + [\ell] Q}. 
\]
Hereafter, the third register is discarded and a quantum Fourier transform ${\rm QFT}_{2^{n+1}}$ on $n+1$ qubits is computed on each of the two registers. Finally, the state of the first two registers---which hold a total of $2(n+1)$ qubits---is measured. As shown in \cite{Sho97}, the discrete logarithm $m$ can be computed from this measurement data via classical post-processing. The corresponding quantum circuit is 
shown in Figure \ref{fig:shorcircuit}. 

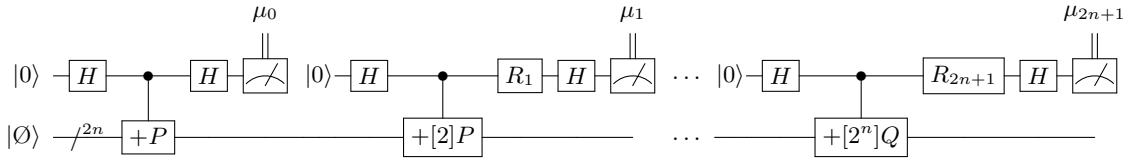
\begin{figure}[h!]
{%\vspace{1em}
\footnotesize
\Qcircuit @C=0.65em @R=1.0em { 
\push{\rule{1.5em}{0em}} & & & & & \ustick{\mu_0}\cwx[1] & & & & & & & & \ustick{\mu_1}\cwx[1] & & & & & & & & & & \ustick{\mu_{2n+1}}\cwx[1] \\
& \lstick{\ket{0}} & \gate{H} & \ctrl{1} & \gate{H} & \meter & & \ket{0} & & \gate{H} & \ctrl{1} & \gate{R_1} & \gate{H} & \meter & & \dots & \push{\rule{0.5em}{0em}} & \ket{0} & & \gate{H} & \ctrl{1} & \gate{R_{2n+1}} & \gate{H} & \meter \\
& \lstick{\ket{\O}} & \qw \slash^{2n} & \gate{+P} & \qw & \qw & \qw & \qw & \qw & \qw & \gate{+[2]P} & \qw  & \qw &  \qw & & \dots & \push{\rule{0.5em}{0em}} & \qw & \qw & \qw & \gate{+[2^n]Q} & \qw &  \qw  & \qw
}
}
\caption{\label{fig:shorcircuit2} Shor's algorithm to compute the discrete logarithm in the subgroup of an elliptic curve generated by a point $P$, analogous to the algorithms from~\cite{Beau03,GN:96,HRS16}. The gates $R_k$ are phase shift gates given by $\mathrm{diag}(1,e^{i\theta_k})$, where 
$\theta_k = -\pi\sum_{j=0}^{k-1}2^{k-j}\mu_j$ and the sum runs over all previous measurements $j$ with outcome $\mu_j\in \{0,1\}$. 
In contrast to the circuit in Figure~\ref{fig:shorcircuit} only one additional qubit is needed besides qubits required to represent and add the elliptic curve points.
}
\end{figure}

Using Kitaev's phase estimation framework \cite{NC:2000}, Beauregard~\cite{Beau03} obtained a quantum algorithm for factoring an integer $N$ from a circuit that performs a conditional multiplication of the form $x \mapsto ax \mod N$, where $a \in \Z_N$ is a random constant integer modulo $N$. The circuit uses only $2n+3$ qubits, where $n$ is the bitlength of the integer to be factored. An implementation of this algorithm on $2n+2$ qubits, using Toffoli-gate-based modular multiplication is described in~\cite{HRS16}. Following the semiclassical Fourier transform method \cite{GN:96}, one can modify Shor's ECDLP algorithm, resulting in the circuit shown in Figure~\ref{fig:shorcircuit2}. The phase shift matrices $R_i = \begin{tiny}\left(
\begin{matrix}
1&0\\
0&e^{i\theta_k}
\end{matrix}\right)
\end{tiny}$, $\theta_k = -\pi\sum_{j=0}^{k-1}2^{k-j}\mu_j$,
depend on all previous measurement outcomes $\mu_j\in \{0,1\}$, $j \in \{0,\dots,k-1\}$. 

\section{Reversible modular arithmetic}
\label{sec:mod}
Shor's algorithm for factoring actually only requires modular multiplication of a quantum integer with classically known constants. In contrast, the elliptic curve discrete logarithm algorithm requires elliptic curve scalar multiplications to compute $[k]P + [\ell]Q$ for a superposition of values for the scalars $k$ and $\ell$. These scalar multiplications are comprised of elliptic curve point additions, which in turn consist of a sequence of modular operations on the coordinates of the elliptic curve points. This requires the implementation of full modular arithmetic, which means that one needs to add and multiply two integers held in quantum registers modulo the constant integer modulus $p$. 

This section presents quantum circuits for reversible modular arithmetic on $n$-bit integers that are held in quantum registers. We provide circuit diagrams for the modular operations, in which black triangles on the right side of gate symbols indicate qubit registers that are modified and hold the result of the computation. Essential tools for implementing modular arithmetic are integer addition and bit shift operations on integers, which we describe first. 

\subsection{Integer addition and binary shifts}
The algorithms for elliptic curve point addition as described below need integer addition and subtraction in different variants: standard integer addition and subtraction of two $n$-bit integers, addition and subtraction of a classical constant integer, as well as controlled versions of those.

For adding two integers, we take the quantum circuit described by Takahashi~et~al.~\cite{TTK10}. The circuit works on two registers holding the input integers, the first of size $n$ qubits and the second of size $n+1$ qubits. It operates in place, i.e. the contents of the second register are replaced to hold the sum of the inputs storing a possible carry bit in the additionally available qubit. To obtain a subtraction circuit, we implement an inverse version of this circuit. The carry bit in this case indicates whether the result of the subtraction is negative. Controlled versions of these circuits can be obtained by using partial reflection symmetry to save controls, which compares favorably to a generic version where simply all gates are controlled. 
For the constant addition circuits, we take the algorithms described in~\cite{HRS16}. Binary doubling and halving circuits are needed for the Montgomery multiplication and inversion algorithms. They are implemented essentially as cyclic bit shifts realized by sequences of symmetric bit swap operations built from $\mathrm{CNOT}$ gates.

\subsection{Modular addition and doubling}
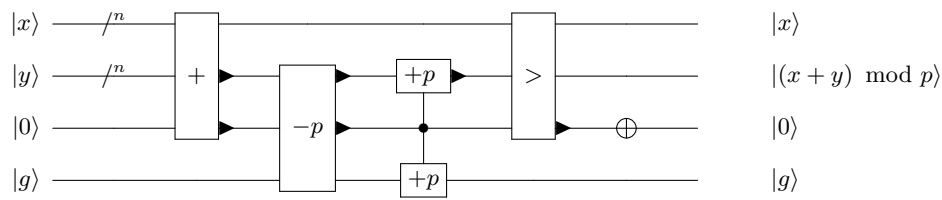
\begin{figure}[b!]
\Qcircuit @C=2.5em @R=1.0em { 
\push{\rule{4em}{0em}} & \lstick{\ket{x}} & \qw \slash^n & \multigateouttwolow{2}{+} & \qw & \qw & \multigateoutlow{2}{>} & \qw & \qw & \rstick{\ket{x}} \\
& \lstick{\ket{y}} & \qw \slash^n & \ghost{+} & \multigateouttwohi{2}{-p} & \gateout{+p} & \ghost{>} & \qw & \qw &\rstick{\ket{(x + y)\!\mod p}}\\
& \lstick{\ket{0}} & \qw  & \ghost{+} & \ghost{-p} & \ctrl{-1} \qwx[1] & \ghost{>} &  \targ & \qw &\rstick{\ket{0}}\\
& \lstick{\ket{g}} & \qw & \qw & \ghost{-p} & \gate{+p} & \qw & \qw & \qw & \rstick{\ket{g}}
} 
\caption{$\mathtt{add\_modp}$: Quantum circuit for in-place modular addition $\ket{x}\ket{y} \mapsto \ket{x}\ket{(x + y)\!\mod p}$. 
The registers $\ket{x}$, $\ket{y}$ consist of $n$ logical qubits each. 
The circuit uses integer addition $+$, addition $+p$ and subtraction $-p$ of the constant modulus $p$, and strict comparison of two $n$-bit integers in the registers $\ket{x}$ and $\ket{y}$, where the output bit flips the carry qubit in the last register. The constant adders use an ancilla qubit in an unknown state $\ket{g}$, which is returned to the same state at the end of the circuit. To implement controlled modular addition $\mathtt{ctrl\_add\_modp}$, one simply controls all operations in this circuit.}\label{fig:modadd}
\end{figure}

We now turn to modular arithmetic. The circuit shown in Figure~\ref{fig:modadd} computes a modular addition of two integers $x$ and $y$ held in $n$-qubit quantum registers $\ket{x}$ and $\ket{y}$, modulo the constant integer modulus $p$. It performs the operation in place $\ket{x}\ket{y} \mapsto \ket{x}\ket{(x+y)\!\mod p}$ and replaces the second input with the result. It uses quantum circuits for plain integer addition and constant addition and subtraction of the modulus. It uses two auxiliary qubits, one of which is used as an ancilla qubit in the constant addition and subtraction and can be in an unknown state to which it will be returned at the end of the circuit. The other qubit stores the bit that determines whether a modular reduction in form of a modulus subtraction actually needs to be performed or not. It is uncomputed at the end by a strict comparison circuit between the result and the first input. Modular subtraction is implemented by reversing the circuit. 

\begin{figure}[t!]
\Qcircuit @C=2.5em @R=1.0em { 
\push{\rule{6em}{0em}} & \lstick{\ket{x_0}} & \qw & \multigateouttwo{2}{\cdot 2} & \multigateouttwohi{3}{-p} & \multigateouttwohi{1}{+p} & \ctrlo{2} & \qw & \rstick{\ket{(2x\!\mod p)_0}} \\
& \lstick{\ket{x_{1,\dots, n-1}}} & \qw \slash^{n-1} & \ghost{\cdot 2} & \ghost{-p} & \ghost{+p} & \qw & \qw & \rstick{\ket{(2x\!\mod p)_{1,\dots,(n-1)}}} \\
& \lstick{\ket{0}} & \qw & \ghost{\cdot 2} & \ghost{-p}  & \ctrl{-1} \qwx[1] & \targ & \qw & \rstick{\ket{0}}\\
& \lstick{\ket{g}} & \qw  & \qw & \ghost{-p} & \gate{+p} & \qw & \qw & \rstick{\ket{g}}
} 
\caption{$\mathtt{dbl\_modp}$: Quantum circuit for in-place modular doubling $\ket{x} \mapsto \ket{2x\!\mod p}$ for an odd constant modulus $p$. 
The registers $\ket{x}$ consists of $n$ logical qubits, the circuit diagram represents the least significant bit separately. 
The circuit uses a binary doubling operation $\cdot 2$ and addition $+p$ and subtraction $-p$ of the constant modulus $p$. The constant adders use an ancilla qubit in an unknown state $\ket{g}$, which is returned to the same state at the end of the circuit.}\label{fig:moddbl}
\end{figure}
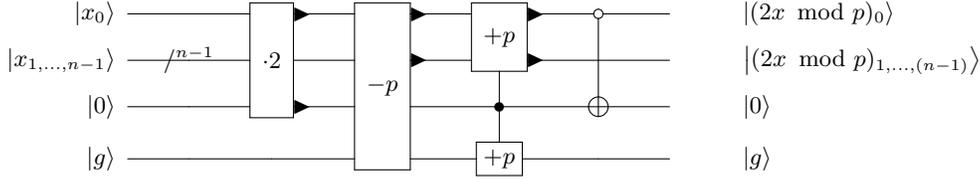

The modular doubling circuit for a constant odd integer modulus $p$ in Figure~\ref{fig:moddbl} follows the same principle. There are two changes that make it more efficient than the addition circuit. First of all it works in place on only one $n$-qubit input integer $\ket{x}$, it computes $\ket{x} \mapsto \ket{2x\!\mod p}$. Therefore it uses only $n+2$ qubits. The first integer addition in the modular adder circuit is replaced by a more efficient multiplication by $2$ implemented via a cyclic bit shift as described in the previous subsection. Since we assume that the modulus $p$ is odd in this circuit, the auxiliary reduction qubit can be uncomputed by checking whether the least significant bit of the result is $0$ or $1$. A subtraction of the modulus has taken place if, and only if this bit is $1$. 

For adding a classical constant to a quantum register modulo a classical constant modulus, we use the in-place modular addition circuit described in~\cite[Section~II]{HRS16}. The circuit operates on the $n$-bit input and requires only 1 ancilla qubit initialized in the state $\ket{0}$ and $n-1$ dirty ancillas that are given in an unknown state and will be returned in the same state at the end of the computation.  

\subsection{Modular multiplication}

\subsubsection{Multiplication by modular doubling and addition.}
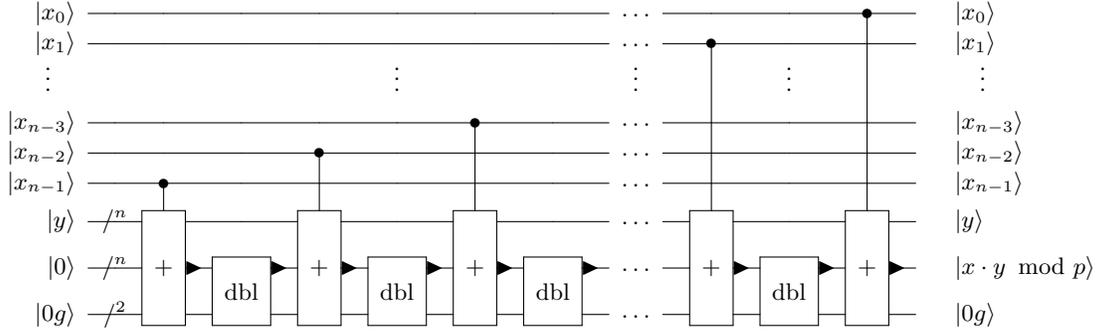
\begin{figure}[h!]
{\footnotesize
\Qcircuit @C=1.1em @R=1.0em { 
\push{\rule{3em}{0em}}& \lstick{\ket{x_{0}}} & \qw & \qw & \qw & \qw & \qw & \qw & \qw & \qw & \dots  & \push{\rule{0em}{0em}} & \qw & \qw & \ctrl{7} & \qw &  \rstick{\ket{x_{0}}} \\
& \lstick{\ket{x_{1}}} & \qw & \qw & \qw & \qw & \qw & \qw & \qw & \qw & \dots & \push{\rule{0em}{0em}} & \ctrl{6} & \qw & \qw & \qw & \rstick{\ket{x_{1}}}\\
& \lstick{\vdots\quad} & & & & & \vdots & & & & \vdots & & & \vdots & & & \rstick{\quad\vdots} \\
& & & & & & & & & & & & & & & & & \\
& \lstick{\ket{x_{n-3}}} & \qw & \qw & \qw & \qw & \qw & \ctrl{3} & \qw & \qw & \dots & \push{\rule{0em}{0em}} & \qw & \qw & \qw & \qw & \rstick{\ket{x_{n-3}}} \\
& \lstick{\ket{x_{n-2}}} & \qw & \qw & \qw & \ctrl{2} & \qw & \qw & \qw & \qw & \dots & \push{\rule{0em}{0em}} & \qw & \qw & \qw & \qw &  \rstick{\ket{x_{n-2}}}\\
& \lstick{\ket{x_{n-1}}} & \qw & \ctrl{1} & \qw & \qw & \qw & \qw & \qw & \qw & \dots  & \push{\rule{0em}{0em}} & \qw & \qw & \qw & \qw & \rstick{\ket{x_{n-1}}}\\
& \lstick{\ket{y}} & \qw \slash^n & \multigateoutsecond{2}{+} & \qw & \multigateoutsecond{2}{+} & \qw & \multigateoutsecond{2}{+} & \qw & \qw & \dots & \push{\rule{0em}{0em}} & \multigateoutsecond{2}{+} & \qw & \multigateoutsecond{2}{+} & \qw & \rstick{\ket{y}}\\
& \lstick{\ket{0}} & \qw \slash^n & \ghost{+} & \multigateout{1}{\dbl} & \ghost{+} & \multigateout{1}{\dbl} & \ghost{+} & \multigateout{1}{\dbl} & \qw & \dots & \push{\rule{0em}{0em}} &  \ghost{+} & \multigateout{1}{\dbl} &  \ghost{+} & \qw & \rstick{\ket{x\cdot y\!\mod p}}\\
& \lstick{\ket{0g}} & \qw \slash^2 & \ghost{+} & \ghost{\dbl} & \ghost{+} & \ghost{\dbl} & \ghost{+} & \ghost{\dbl} & \qw & \dots & \push{\rule{0em}{0em}} &  \ghost{+} & \ghost{\dbl} &  \ghost{+} & \qw & \rstick{\ket{0g}}
}
}
\caption{$\mathtt{mul\_modp}$: Quantum circuit for modular multiplication $\ket{x}\ket{y}\ket{0} \mapsto \ket{x}\ket{y}\ket{x\cdot y\!\mod p}$ built from modular doublings $\dbl \gets \mathtt{dbl\_modp}$ and controlled modular additions $+ \gets \mathtt{ctrl\_add\_modp}$. 
The registers $\ket{x_i}$ hold single logical qubits, $\ket{y}$ and $\ket{0}$ hold $n$ logical qubits. The two ancilla qubits $\ket{0g}$ are the ones needed in the modular addition and doubling circuits, the second one can be in an unknown state to which it will be returned.
}\label{fig:modmulDBLADD}
\end{figure}

Modular multiplication can be computed by repeated modular doublings and conditional modular additions. Figure~\ref{fig:modmulDBLADD} shows a circuit that computes the product $z=x\cdot y\!\mod p$ for constant modulus $p$ as described by Proos and Zalka~\cite[Section~4.3.2]{PZ03} by using a simple expansion of the product along a binary decomposition of the first multiplicand, i.e.
\[x\cdot y = \sum_{i=0}^{n-1}x_i2^i \cdot y =  x_0y + 2(x_1y + 2(x_2y +   \dots + 2(x_{n-2}y+2(x_{n-1}y))\dots)). \]
The circuit runs on $3n+2$ qubits, $2n$ of which are used to store the inputs, $n$ to accumulate the result and $2$ ancilla qubits are needed for the modular addition and doubling operations, one of which can be dirty. The latter could be taken to be one of the $x_i$, for example $x_0$ except in the last step, when the modular addition gate is conditioned on $x_0$. For simplicity, we assume it to be a separate qubit. 

\begin{figure}[h!]
{\footnotesize
\Qcircuit @C=1.2em @R=1.0em { 
\push{\rule{4em}{0em}} & \lstick{\ket{0}} & \targ & \ctrl{1} & \targ & \targ & \ctrl{1} & \targ & \qw & \dots & \push{\rule{0em}{0em}} & \targ & \targ & \ctrl{1} & \targ & \qw & \rstick{\ket{0}} \\ 
& \lstick{\ket{x_0}} & \qw & \multigate{8}{+} & \qw & \qw & \multigate{8}{+} & \qw & \qw & \dots  & \push{\rule{0em}{0em}} & \qw & \ctrl{-1} & \multigate{8}{+} & \ctrl{-1} & \qw & \rstick{\ket{x_0}}\\
& \lstick{\ket{x_1}} & \qw & \ghost{+} & \qw & \qw & \ghost{+} & \qw & \qw & \dots & \push{\rule{0em}{0em}} & \ctrl{-2} & \qw & \ghost{+} & \qw & \qw &  \rstick{\ket{x_1}}\\
& \lstick{\vdots\quad} & & & \vdots & & & & & \vdots & & & & & & \rstick{\quad\vdots} \\
& & & & & & & & & & & & & & & & \\
& \lstick{\ket{x_{n-3}}} & \qw & \ghost{+} & \qw & \qw & \ghost{+} & \qw & \qw & \dots & \push{\rule{0em}{0em}} & \qw & \qw & \ghost{+} & \qw & \qw & \rstick{\ket{x_{n-3}}} \\
& \lstick{\ket{x_{n-2}}} & \qw & \ghost{+} & \qw & \ctrl{-6} & \ghost{+} & \ctrl{-6} & \qw & \dots & \push{\rule{0em}{0em}} & \qw & \qw & \ghost{+} & \qw  & \qw & \rstick{\ket{x_{n-2}}}\\
& \lstick{\ket{x_{n-1}}} & \ctrl{-7} & \ghost{+} & \ctrl{-7} & \qw &  \ghost{+} & \qw & \qw & \dots  & \push{\rule{0em}{0em}} & \qw & \qw &  \ghost{+} & \qw & \qw &  \rstick{\ket{x_{n-1}}} \\
& \lstick{\ket{0}} & \qw \slash^n & \ghost{+} & \multigateout{1}{\dbl} & \qw & \ghost{+}  & \multigateout{1}{\dbl} & \qw & \dots & \push{\rule{0em}{0em}} & \multigateout{1}{\dbl} & \qw &  \ghost{+} & \qw & \qw & \rstick{\ket{x^2\!\mod p}}\\
& \lstick{\ket{0g}} & \qw \slash^2 & \ghost{+} & \ghost{\dbl} & \qw & \ghost{+} & \ghost{\dbl} & \qw & \dots & \push{\rule{0em}{0em}} &  \ghost{\dbl} & \qw &  \ghost{+} & \qw & \qw & \rstick{\ket{0g}}
}
}
\caption{$\mathtt{squ\_modp}$: Quantum circuit for modular squaring $\ket{x}\ket{0} \mapsto \ket{x}\ket{x^2\!\mod p}$ built from modular doublings $\dbl \gets \mathtt{dbl\_modp}$ and controlled modular additions $+ \gets \mathtt{ctrl\_add\_modp}$. 
The registers $\ket{x_i}$ hold single logical qubits, $\ket{0}$ holds $n$ logical qubits. The two ancilla qubits $\ket{0g}$ are the ones needed in the modular addition and doubling circuits, the second one can be in an unknown state to which it will be returned.
}\label{fig:modsquDBLADD}
\end{figure}
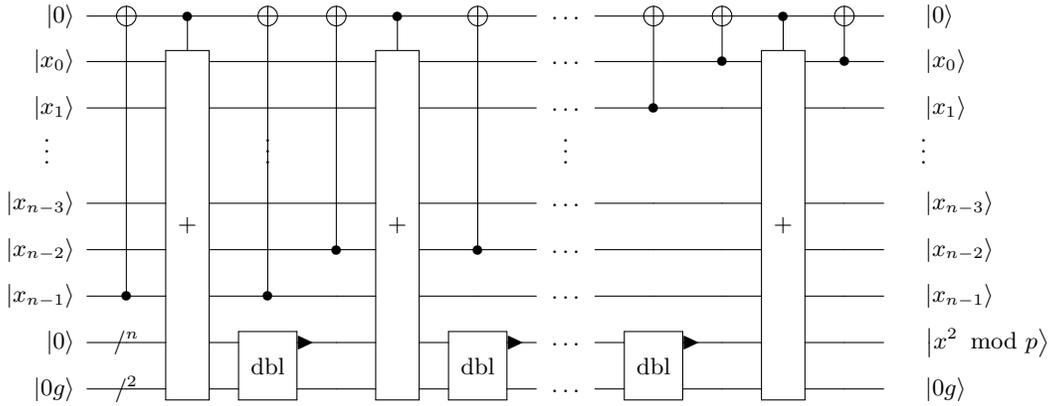

Figure~\ref{fig:modsquDBLADD} shows the corresponding specialization to compute a square $z=x^2\!\mod p$. It uses $2n+3$ qubits by removing the $n$ qubits for the second input multiplicand, and adding one ancilla qubit, which is used in round $i$ to copy out the current bit $x_i$ of the input in order to add $x$ to the accumulator conditioned on the value of $x_i$.

\subsubsection{Montgomery multiplication}
In classical applications, Montgomery multiplication~\cite{mont85} is often the most efficient choice for modular multiplication if the modulus does not have a special shape such as being close to a power of $2$. Here, we explore it as an alternative to the algorithm using modular doubling and addition as described above. 

In~\cite{mont85}, Montgomery introduced a representation for an integer modulo $p$ he called a $p$-residue that is now called the Montgomery representation. Let $R>p$ be an integer radix coprime to $p$. An integer $a$ modulo $p$ is represented by the Montgomery representation $aR\!\mod p$. The Montgomery reduction algorithm takes as input an integer $0 \leq c < Rp$ and computes $cR^{-1}\!\mod p$. Thus given two integers $aR\!\mod p$ and $bR\!\mod p$ in Montgomery representation, applying the Montgomery reduction to their product yields the Montgomery representation $(ab)R\!\mod p$ of the product. If $R$ is a power of $2$, one can interleave the Montgomery reduction with school-book multiplication, obtaining a combined Montgomery multiplication algorithm. The division operations usually needed for computing remainders are replaced by binary shifts in each round of the multiplication algorithm.

\begin{figure}[t!]
{\footnotesize
\Qcircuit @C=0.6em @R=0.9em { 
\push{\rule{3em}{0em}}
& \lstick{\ket{x_0}} & \qw & \ctrl{5} & \qw & \qw & \qw & \qw & \qw & \qw & \qw & \qw & \dots  & \push{\rule{0em}{0em}} & \qw & \qw & \qw & \qw & \qw & \qw & \qw & \rstick{\ket{x_0}}\\
& \lstick{\ket{x_1}} & \qw & \qw & \qw & \qw & \qw & \ctrl{4} & \qw & \qw & \qw & \qw &  \dots & \push{\rule{0em}{0em}} & \qw & \qw & \qw & \qw & \qw & \qw & \qw & \rstick{\ket{x_1}}\\
& \lstick{\vdots\quad} & & & & & \vdots & &  & & & & \vdots & & & \vdots & & & & & & \rstick{\quad\vdots} \\
& & & & & & & & & & & & & & & & & & & & & & \\
& \lstick{\ket{x_{n-1}}} & \qw &\qw& \qw & \qw & \qw & \qw & \qw & \qw & \qw & \qw & \dots  & \push{\rule{0em}{0em}} & \ctrl{1} & \qw & \qw & \qw & \qw & \qw & \qw &  \rstick{\ket{x_{n-1}}} \\
& \lstick{\ket{y}} & \qw\slash & \multigateoutthreelow{3}{+} & \qw & \qw & \qw & \multigateoutthreelow{3}{+} & \qw & \qw & \qw & \qw & \dots  & \push{\rule{0em}{0em}} & \multigateoutthreelow{3}{+} & \qw & \qw & \qw & \qw & \qw & \qw & \rstick{\ket{y}}\\
& \lstick{\ket{0}} & \qw & \ghost{+} & \qw & \multigateoutthreeup{3}{+p} &  \multigateoutthreeup{2}{/2} &  \ghost{+} & \qw & \multigateoutthreeup{3}{+p} &  \multigateoutthreeup{2}{/2} & \qw & \dots  & \push{\rule{0em}{0em}} & \ghost{+} & \qw & \multigateoutthreeup{3}{+p} &\multigateoutthreeup{2}{/2} & \multigateoutthreeup{3}{-p} & \ctrl{1} & \qw & \rstick{\ket{c}} \\
& \lstick{\ket{0..0}} & \qw\slash & \ghost{+} & \qw & \ghost{+p} & \ghost{/2} &  \ghost{+} & \qw & \ghost{+p} & \ghost{/2} & \qw & \dots  & \push{\rule{0em}{0em}} & \ghost{+} & \qw & \ghost{+p} & \ghost{/2} & \ghost{-p} & \multigateouttwohi{2}{+p} & \qw & \rstick{\ket{z_{1,\dots,n-1}}}\\
& \lstick{\ket{0}} & \qw & \ghost{+} & \ctrl{2} & \ghost{+p} &  \ghost{/2} & \ghost{+} & \ctrl{3} & \ghost{+p} & \ghost{/2} & \qw & \dots  & \push{\rule{0em}{0em}} & \ghost{+} & \ctrl{6} &\ghost{+p} &  \ghost{/2} & \ghost{-p} & \ghost{+p} & \qw & \rstick{\ket{z_0}}\\
& \lstick{\ket{g}} & \qw & \qw & \qw & \ghost{+p} & \qw & \qw & \qw & \ghost{+p} & \qw & \qw & \dots  & \push{\rule{0em}{0em}} & \qw & \qw & \ghost{+p} & \qw & \ghost{-p} & \ghost{+p} & \qw & \rstick{\ket{g}}\\
& \lstick{\ket{0}} & \qw & \qw & \targ & \ctrl{-1} & \qw & \qw & \qw & \qw & \qw & \qw & \dots  & \push{\rule{0em}{0em}} & \qw & \qw & \qw & \qw & \qw & \qw & \qw & \rstick{\ket{m_0}}\\
& \lstick{\ket{0}} & \qw & \qw & \qw & \qw & \qw & \qw & \targ & \ctrl{-2} & \qw & \qw & \dots & \push{\rule{0em}{0em}} & \qw & \qw & \qw & \qw & \qw & \qw & \qw & \rstick{\ket{m_1}}\\
& \lstick{\vdots\,} & & & & & \vdots & & & & & & \vdots & & & & & & \vdots & & & \rstick{\quad\vdots} \\
& & & & & & & & & & & & & & & & & & & & & &\\
& \lstick{\ket{0}} & \qw & \qw & \qw & \qw & \qw & \qw & \qw & \qw & \qw & \qw & \dots & \push{\rule{0em}{0em}} & \qw & \targ & \ctrl{-5} & \qw & \qw & \qw & \qw & \rstick{\ket{m_{n-1}}}
}
}
\caption{$\mathtt{mul\_modp}$: Quantum circuit for the forward Montgomery modular multiplication $\ket{x}\ket{y}\ket{0} \mapsto \ket{x}\ket{y}\ket{z=x\cdot y\cdot R^{-1}\!\mod p}$. 
The register $\ket{y}$ holds $n$ logical qubits and $\ket{0..0}$ holds $n-1$. All others are single qubits. 
The qubit $\ket{g}$ is a dirty ancilla qubit in an unknown state. The qubit labeled $\ket{m_i}$ holds the information whether the intermediate result in round $i$ was odd and thus whether $p$ was added to it. The circuit uses integer addition $+$, integer addition $+p$ and subtraction $-p$ of the constant modulus $p$ and a halving circuit $/2$ that performs a cyclic qubit shift. The last two gates reflect the standard conditional subtraction of $p$. To uncompute the qubits $\ket{m_i}$, one copies out the result $z$ and runs the circuit backwards.}\label{fig:modmulMONT}
\end{figure}
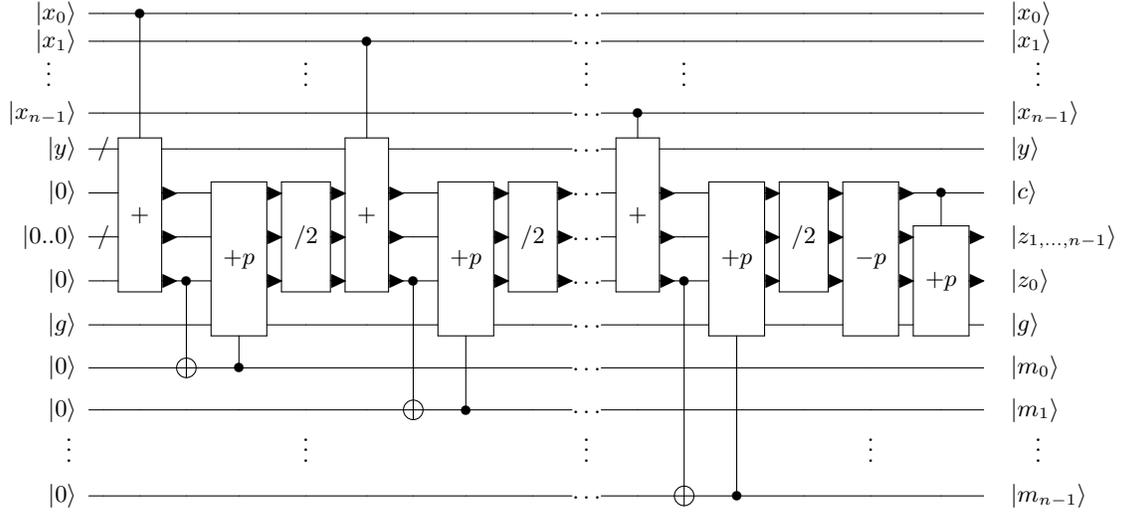

The multiplication circuit using modular doubling and addition operations described in the previous subsection contains two modular reductions in each round of the algorithm. Each of those is realized here by at least two integer additions. In contrast, the Montgomery algorithm shown in Figure~\ref{fig:modmulMONT} avoids these and uses only one integer addition per round. This reduces the circuit depth in comparison to the double-and-add approach. However, it comes at the cost of requiring more qubits. The main issue is that the algorithm stores the information for each round, whether the odd modulus $p$ had to be added to the intermediate result to make it even or not. This is done to allow divisions by $2$ through a simple bit shift of an even number. These bits are still set at the end of the circuit shown in 
Figure~\ref{fig:modmulMONT}. To uncompute these values, we copy the result to another $n$-qubit register, and run the algorithm backwards, which essentially doubles the depth of the algorithm. But this still leads to a lower overall depth than the one of the double-and-add algorithm. Hence, switching to Montgomery multiplication presents a trade-off between the required number of qubits and the multiplication circuit depth. 

The same optimization as shown in the previous section allows to save $n-1$ qubits when implementing a Montgomery squaring circuit that computes $z=x^2\!\mod p$.

\subsection{Modular inversion}\label{sec:modinv}
Performing the modular inversion on a quantum computer is by far the most costly operation required in order to implement the affine group law of an elliptic curve. We use a reversible circuit for the extended binary greatest common divisor algorithm~\cite{St67} that implements Kaliski's algorithm~\cite{Kal95} for inverting a number $x R\!\mod\!p$ given in Montgomery representation for $R=2^n$; i.e. an algorithm (i) which only uses elementary reversible operations such as Toffoli gates, (ii)  whose sequence of instructions does not depend on the given input $x 2^n\!\mod\!p$, and (iii) whose output is again in Montgomery form $x^{-1}2^n\!\mod\!p$.  

We use the extended binary GCD algorithm to compute the representation of the greatest common divisor between $x$ and $p$, with the added requirement to ensure property (ii), namely to make sure that the sequence of operations that carries out the Euclidean algorithm is the same, independent of $x$. In particular, an issue is that for different inputs $x\not= x'$ the usual, irreversible Euclidean algorithm can terminate in a different number of steps. To fix this, we include a counter register which is incremented upon termination of the loop to ensure the algorithm is run for $2n$ rounds, which is the worst-case runtime. 

In the following algorithm to compute the Montgomery inverse the inputs are a prime $p$ and a value $x$ where $0 \leq x < p$. The output is $x^{-1} 2^n \; {\rm mod} \; p$. In functional programming style (here, using F\# syntax), Kaliski's algorithm is described as follows: 

\newpage
{
\footnotesize
\begin{lstlisting}[language=FSharp]
let MGinverse p x = 
  let rec xmg u v r s k =
    match u, v, r, s with 
      | _,0,r,_             -> r
      | u,_,_,_ when u\%2=0 -> xmg (u >>> 1) v r (s <<< 1) (k+1) 
      | _,v,_,_ when v\%2=0 -> xmg u (v >>> 1) (r <<< 1) s (k+1) 
      | u,v,_,d when u > v  -> xmg ((u-v) >>> 1) v (r+s) (s <<< 1) (k+1) 
      | _,_,_,_             -> xmg u ((v-u) >>> 1) (r <<< 1) (r+s) (k+1) 
  xmg p x 0 1 0 
\end{lstlisting}
}
The algorithm actually computes only the so-called ``almost inverse'' which is of the form $x^{-1} 2^k$, i.e., there is a secondary step necessary to convert to the correct form (not shown here). Note that here $k$ depends on $x$, i.e., it is necessary to uncompute $k$. Two example executions are shown in Figure \ref{fig:kaliski}. 

\begin{figure}
\centering
\begin{tabular}{c@{\qquad\qquad}c}
\begin{tabular}{m{4mm}m{4mm}m{4mm}m{4mm}m{4mm}m{4mm}m{4mm}m{4mm}m{4mm}m{4mm}}
\hline\hline
$u$    & 11 & 11 & 11 & 11 & 5 & 2 & 1 & 1  & 1 \\
$v$    & 8  &  4 &  2 &  1 & 1 & 1 & 1 & 0  & 0 \\
$r$    & 0  &  0 &  0 &  0 & 1 & 3 & 3 & 6  & 6 \\
$s$    & 1  &  1 &  1 &  1 & 2 & 4 & 8 & 11 & 11\\
$k$    & 0  &  1 &  2 &  3 & 4 & 5 & 6 & 7  & 7 \\
$\ell$ & 0  &  0 &  0 &  0 & 0 & 0 & 0 & 0  & 1 \\
\hline\hline
\end{tabular} & 
\begin{tabular}{m{4mm}m{4mm}m{4mm}m{4mm}m{4mm}m{4mm}m{4mm}m{4mm}m{4mm}m{4mm}}
\hline\hline
$u$    & 11 &  2 &  1 &  1 & 1 & 1 & 1 & 1  & 1 \\
$v$    & 7  &  7 &  7 &  3 & 1 & 0 & 0 & 0  & 0 \\
$r$    & 0  &  1 &  1 &  2 & 4 & 8 & 8 & 8  & 8 \\
$s$    & 1  &  2 &  4 &  5 & 7 & 11 & 11 & 11 & 11\\
$k$    & 0  &  1 &  2 &  3 & 4 & 5 & 5 & 5  & 5 \\
$\ell$ & 0  &  0 &  0 &  0 & 0 & 0 & 1 & 2  & 3 \\
\hline\hline
\end{tabular}\\
\raisebox{-0.2cm}{(a)} & \raisebox{-0.2cm}{(b)} 
\end{tabular}
\caption{\label{fig:kaliski} Two example runs of the reversible extended binary Euclidean algorithm to compute the Montgomery inverse modulo $p=11$. Shown in (a) is the execution for input $x=8$ which leads to termination of the usual irreversible algorithm after $k=7$ steps. The algorithm is always executed for $2n$ rounds, where $n$ is the bit-size of $p$ which is an upper bound on the maximum number of steps required for general input $x$. Once the final step $v=0$ has been reached, a counter register $\ell$ is incremented. Shown in (b) is the execution for input $x=7$ which leads to termination after $5$ steps after which the counter is incremented three times.}
\end{figure}

As shown in Fig.~\ref{fig:kaliski}, the actual number of steps that need to be executed until the gcd is obtained, depends on the actual input $x$: in the first example the usual Kaliski algorithm terminates after $k=7$ steps, whereas in the second example the usual algorithm would terminate after $k=5$ steps. 
To make the algorithm reversible, we must find an implementation that carries out the same operations, irrespective of the input. The two main ingredients to obtain such an implementation are a) an upper bound of $2n$ steps that Kaliski's algorithm can take in the worst case \cite{Kal95} and b) the introduction of a counter that ensures that either the computation is propagated forward or, in case the usual Kaliski algorithm has terminated, the counter is incremented.  Such a counter can be implemented using $O(\log(n))$ qubits. 

\begin{figure}[hbt!]
{\footnotesize
\Qcircuit @C=0.85em @R=1.0em { 
\push{\rule{2em}{0em}}& \lstick{\ket{u}}  &  \qw\slash^n & \qw & \qw & \qw & \ctrlo{6} & \qw & \multigateouttwolow{2}{-} & \qw & \qw & \qw& \multigateouttwolow{2}{+} & \gate{/2}\qwx[3] & \qw & \multigateout{1}{-} & \gate{/2} \qwx[3] & \multigateoutsecond{1}{-} & \qw & \qw & \qw & \qw &  \rstick{\ket{u}}\\
& \lstick{\ket{v}} &  \qw\slash^n  & \qw & \ctrlo{8}& \qw & \qw & \ctrlo{5} & \ghost{-} & \qw & \qw & \qw & \ghost{+} & \qw & \gate{/2}\qwx[3] & \ghost{-}\qwx[2] & \qw & \ghost{-}\qwx[2] & \gate{/2}\qwx[3] & \qw & \qw & \qw & \rstick{\ket{v}} \\
& \lstick{\ket{0}} & \qw & \qw & \qw & \qw & \qw & \qw & \ghost{-} & \qw & \ctrl{3}  & \qw & \ghost{+}& \qw & \qw & \qw & \qw & \qw & \qw & \qw & \qw & \qw & \rstick{\ket{0}}  \\
& \lstick{\ket{s}} & \qw\slash^n & \qw & \qw & \qw & \qw & \qw & \qw & \qw & \qw & \qw & \qw & \gate{\cdot 2} & \qw & \multigateoutlow{1}{+} & \gate{\cdot 2} & \multigateout{1}{+} & \qw & \qw & \qw & \qw& \rstick{\ket{s}} \\
& \lstick{\ket{r}} & \qw\slash^n & \qw & \qw & \qw & \qw & \qw & \qw & \qw & \qw & \qw & \qw & \qw & \gate{\cdot 2} & \ghost{+} & \qw & \ghost{+} & \gate{\cdot 2} & \ctrl{2} & \qw & \qw& \rstick{\ket{r}} \\
& \lstick{\ket{0}} & \qw & \qw & \qw & \qw & \qw & \qw & \targ & \targ & \ctrlo{1} & \targ & \targ & \qw & \qw  & \qw & \qw & \qw & \qw & \qw & \qw & \qw & \rstick{\ket{0}} \\
& \lstick{\ket{0}} & \qw & \qw & \qw & \qw & \targ & \ctrlo{1} & \ctrl{-1} & \qw & \targ\qwx[1] & \qw & \ctrl{-1} & \ctrl{-3} & \ctrlo{-2} & \ctrl{-2} & \ctrl{-3} & \ctrlo{-2} & \ctrlo{-2} & \targ & \qw & \qw & \rstick{\ket{0}}  \\
& \lstick{\ket{m_i}} & \qw & \qw & \qw & \qw & \qw & \targ & \qw & \ctrl{-2} & \targ & \ctrl{-2} & \qw & \ctrlo{-1} & \ctrl{-1} & \ctrl{-1} & \ctrl{-1} & \ctrlo{-1} & \ctrlo{-1} & \qw & \qw & \qw & \rstick{\ket{m_i}} \\
& & & & & & & & & & & & & & & & & & & & & & & &  \\
& \lstick{\ket{f}} & \qw & \targ & \targ & \targ & \qw & \qw & \qw & \qw & \qw & \ctrlo{1} & \qw & \ctrl{-1} & \qw & \qw & \qw & \qw & \qw & \qw & \qw & \qw& \rstick{\ket{f}} \\
& \lstick{\ket{k}}  & \qw\slash^{l} & \ctrl{-1}& \qw & \qw & \qw & \qw & \qw & \qw & \qw & \gate{\mathrm{INC}} & \qw & \qw & \qw & \qw & \qw & \qw & \qw & \qw & \qw & \qw & \rstick{\ket{k}}  
\gategroup{1}{7}{8}{20}{2.2em}{--}
}
}
\caption{Quantum circuit for the Montgomery-Kaliski round function. The function is repeated $2n$ times to obtain a reversible modular inversion algorithm. The $n$-qubit registers $\ket{u}$, $\ket{v}$, $\ket{r}$, $\ket{s}$ represent the usual coefficients in the binary Euclidean algorithm. The circuit uses integer subtraction $-$ and addition $+$, as well as multiplication and division by $2$ functions $\cdot 2$ and $/2$ and an incrementer circuit $\mathrm{INC}$. The circuits $\cdot 2$ and $/2$ are implemented as cyclic qubit shifts.}\label{fig:modinv}
\end{figure}
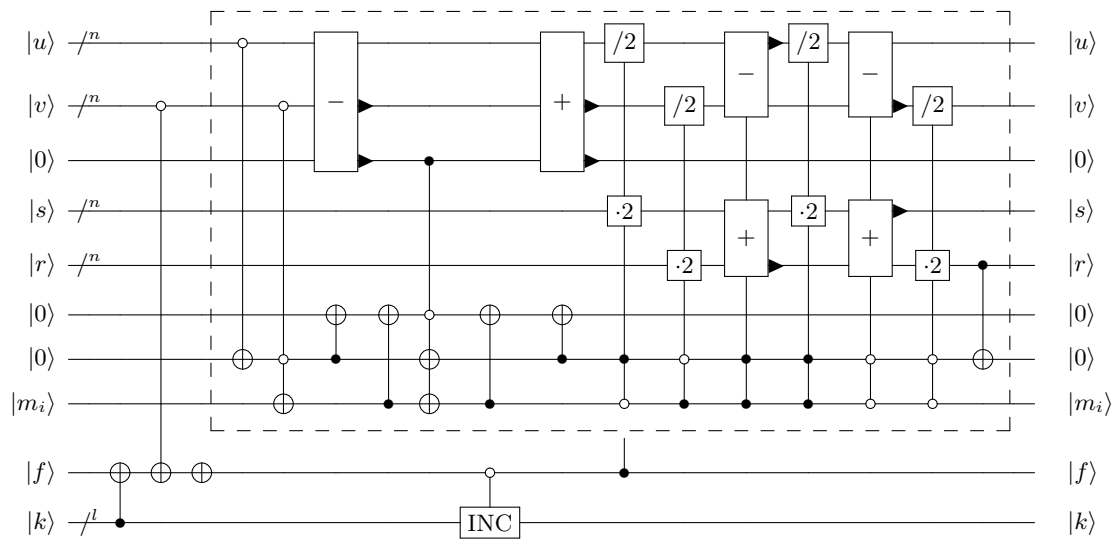

The circuit shown in Fig.~\ref{fig:modinv} implements the Kaliski algorithm in a reversible fashion. We next describe the various registers used in this circuit and explain why this algorithm actually computes the same output as the Kaliski algorithm. The algorithm uses $n$-bit registers for inputs $u$ and $v$, where $u$ is  initially set to the underlying prime $p$. As $p$ is constant, the register can be prepared using bit flips corresponding to the binary representation of $p$. Moreover, $v$ is initially set to the input $x$ of which we would like to compute the inverse. 
Moving downward from the top, the next line represents a single ancilla qubit which is used to store an intermediate value which is the result of a comparison. Next is an $n+1$-bit register for $r$ and likewise an $n+1$-bit register for $s$, so that the loop invariant $p=rv+su$ holds at each stage of the algorithm. Eventually, when $v=0$ is reached, register $r$ will hold the almost inverse and register $s$ will be equal to $p$. The next $2$ lines represent ancilla qubits which are used as scratch space to store an intermediate computation. The technically most interesting part is the next register which consists of a single qubit labeled $m_i$. This indicates that in round $i$, where $1\leq i \leq 2n$, a fresh qubit is introduced, then acted upon by the circuit and then kept around. 

After the maximum number of $2n$ rounds is executed, hence $2n$ qubits have been introduced and entangled in this way. The purpose of the qubit $m_i$ is to remember which of the $4$ branches in Kaliski's algorithm was taken in step $i$. As there are $4$ branches, this choice could be naively encoded into $2$ qubits, which however would lead to a space overhead of $4n$ instead of $2n$. The fact that one of these two qubits is actually redundant is shown below. The next qubit, labeled $f$ in the figure, is part of a mechanism to unroll the entire algorithm which drives precisely one of two processes forward: either the Kaliski algorithm itself, or a counter, here represented as the ``INC'' operation. The flag $f$ starts out in state $1$ which indicates that the algorithm is in Kaliski-mode. Once the terminating condition $v=0$ is reached, the flag switches to $0$, indicating that the algorithm is in counter-mode. Finally, the register $k$ holds the state of the counter. As the counter can take values between $n$ and $2n$ only \cite{Kal95}, it can be implemented using $\lceil \log_2(n)+1 \rceil$ qubits only. 

Having covered all registers that are part of the circuit, we next explain how the circuit is actually unraveled to compute the almost inverse. Shown in Fig.~\ref{fig:modinv} is only one round. The circuit is applied over and over to the same set of qubit registers, with the sole exception of qubit $m_i$ which depends on round $i$ and which is initialized, acted upon, and then stored. 
In each round there are $4$ possible branches. These are dispatched using the gates inside the dashed box. The first gate is a controlled NOT that acts only on the least significant bit of $u$, checking whether $u$ is even. The next gate does the same for $v$, which flips the target bit in case $u$ was odd and $v$ was even. If both $u$ and $v$ are odd, the difference $u-v$ respectively $v-u$ has to be computed, depending on whether $u>v$ or $u \leq v$. To figure out which case actually holds, we use a subtractor and store the most significant qubit in the mentioned ancilla. The sequences of $5$ gates underneath the two subtractors/adders serve as an encoder that prepares the following correspondence: `10' for the case $u$ even, '01' for the case $u$ odd, $v$ even, '11' for the case both odd and $u > v$, and '00' for the case both odd and $u \leq v$. Denote the two bits involved in this encoding as '$ab$' we see that $b$ is the round qubit $m_i$. The fact that $a$ can be immediately uncomputed is a consequence of the following observation.

In each step of Kaliski's algorithm, precisely one of $r$ and $s$ is even and the other is odd. If the updated value of $r$ is even, then the branch must be the result of either the case $v$ even or the case both $u$ and $v$ odd and $u \leq v$. Correspondingly, if the updated value of $s$ is even, then the branch must have been the result of either the case $u$ even or the case both $u$ and $v$ odd and $u > v$. Indeed, an even value of $r$ arises only from the mentioned two branches $v$ even or $u$ and $v$ both odd and $u \leq v$. Similarly, the other statement is obtained for $s$. The invariant $p=rv+su$ implies inductively that precisely one of $r$ and $s$ is even and the other henceforth must be odd. 

Coming back to the dashed circuit, the next block of $6$ gates is to dispatch the appropriate case, depending on the $2$ bits $a$ and $b$, which corresponds to the $4$ branches in the match statement. Finally, the last CNOT gate between the least significant bit of $r$ (indicating whether $r$ is even) is used to uncompute 'a'. 

The shown circuit is then applied precisely $2n$ times. At this juncture, the computation of the almost inverse will have stopped after $k$ steps where $n \leq k \leq 2n$ and the counter INC will have been advanced precisely $2n-k$ times. The counter INC could be implemented using a simple increment $x \mapsto x+1$, however in  our implementation we chose a finite state machine that has a transition function requiring less Toffoli gates. 

Next, the register $r$ which is known to hold $-x^{-1} 2^k$ is converted to $x^{-1} 2^n$. This is done by performing precisely $n-k$ controlled modular doublings and a sign flip. Finally, the result is copied out into another register and the entire circuit is run backwards.

\section{Reversible elliptic curve operations}
\label{sec:ecadd}

Based on the reversible algorithms for modular arithmetic from the previous section, we now turn to implementing a reversible algorithm for adding two points on an elliptic curve. Next, we describe a reversible point addition in the generic case in which none of the exceptional cases of the simple affine Weierstrass group law occurs. After that, we describe a reversible algorithm for computing a scalar multiplication $[m]P$.
  
\subsection{Point addition}
The reversible point addition we implement is very similar to the one described in Section~4.3 of \cite{PZ03}. It uses affine coordinates. As was also mentioned in \cite{PZ03} (and as we argue below), it is enough to consider the generic case of an addition. This means that we assume the following situation. Let $P_1, P_2 \in E(\F_p)$, $P_1, P_2 \neq \O$ such that $P_1 = (x_1, y_1)$ and $P_2 = (x_2, y_2)$. Furthermore let, $x_1 \neq x_2$ which means that $P_1 \neq \pm P_2$. Recall that then $P_3 = P_1 + P_2 \neq \O$ and it is given by $P_3 = (x_3, y_3)$, where $x_3 = \lambda^2 - x_1 - x_2$ and $y_3 = \lambda(x_1-x_3) - y_1$ for $\lambda = (y_1 - y_2)/(x_1-x_2)$.

As explained in~\cite{PZ03}, for computing the sum $P_3$ reversibly and in place (replacing the input point $P_1$ by the sum), the algorithm makes essential use of the fact that the slope $\lambda$ can be re-computed from the result $P_3$ via the point addition $P_3 + (-P_2)$ independent of $P_1$ using the equation
\[
\frac{y_1-y_2}{x_1-x_2} = -\frac{y_3 + y_2}{x_3 - x_2}.
\]
Algorithm~\ref{alg:ECpointadd} depicts our algorithm for computing a controlled point addition. As input it takes the four point coordinates for $P_1$ and $P_2$, a control bit $\ctr$, and replaces the coordinates holding $P_1$ with the result $P_3 = (x_3, y_3)$. Note that we assume $P_2$ to be a constant point that has been classically precomputed, because we compute scalar multiples of the input points $P$ and $Q$ to Shor's algorithm by conditionally adding together precomputed $2$-power multiples of these points as shown in Figure~\ref{fig:shorcircuit} above. The point $P_2$ will thus always be one of these values. Therefore, operations involving the coordinates $x_2$ and $y_2$ are implemented as constant operations.
Algorithm~\ref{alg:ECpointadd} uses two additional temporary variables $\lambda$ and $t_0$. All the point coordinates and the temporary variables fit in $n$-bit registers and thus the algorithm can be implemented with a circuit on a quantum register $|x_1\ y_1\ \ctr\ \lambda\ t_0\ \tmp\rangle$, where the register $\tmp$ holds auxiliary registers that are needed by the modular arithmetic operations used in Algorithm~\ref{alg:ECpointadd} as described in Section~\ref{sec:mod}.

The algorithm is given as a straight line program of (controlled) arithmetic operations on the point coefficients and auxiliary variables. The comments at the end of the line after each operation show the current values held in the variable that is possibly changed. The notation $[\cdot]_1$ shows the value of the variable in case the control bit is $\ctr = 1$, if it is $\ctr = 0$ instead, the value is shown with $[\cdot]_0$. In the latter case, it is easy to check that the algorithm indeed returns the original state of the register.

The functions in the algorithm all use the fact that the modulus $p$ is known as a classical constant. They relate to the algorithms described in Section~\ref{sec:mod} as follows: 
\begin{itemize}
\item \texttt{add\_const\_modp} is a modular addition of a constant to a quantum state, \texttt{sub\_const\_modp} is its reverse, a modular subtraction of a constant. 
\item \texttt{ctrl\_add\_const\_modp} is single qubit controlled modular addition of a constant to a qubit register, i.e. the controlled version of the above. Its reverse \texttt{ctrl\_sub\_const\_modp} performs the controlled modular subtraction.
\item \texttt{ctrl\_sub\_modp} is a single qubit controlled modular subtraction on two qubit registers, implemented as the reverse of the corresponding modular addition.
\item \texttt{ctrl\_neg\_modp} is a single qubit controlled modular negation on a register.
\item \texttt{mul\_modp}, \texttt{squ\_modp}, \texttt{inv\_modp} are the out-of-place modular multiplication, squaring and inversion algorithms on two input qubit registers, respectively.
\end{itemize}

\begin{algorithm}[h!]
\caption{Reversible, controlled elliptic curve point addition. This algorithm operates on a quantum register holding the point $P_1 = (x_1,y_1)$, a control bit $\ctr$, and two auxiliary values $\lambda$ and $t_0$. In addition it needs auxiliary registers for the functions that are called as described for those functions.  The second point $P_2 = (x_2, y_2)$ is assumed to be a precomputed classical constant. For $P_1, P_2 \neq \O$, $P_1 \neq \pm P_2$, if $\ctr = 1$, the algorithm correctly computes $P_1\gets P_1 + P_2$ in the register holding $P_1$; if $\ctr = 0$ it returns the register in the received state.}\label{alg:ECpointadd}
\setlength{\columnsep}{1cm}
\begin{multicols}{2}
\begin{algorithmic}[1]
\State \texttt{sub\_const\_modp} $x_1$ $x_2$;             % // x1 <- x1 - x2
\State \texttt{ctrl\_sub\_const\_modp} $y_1$ $y_2$ $\ctr$;   % // y1 <- y1 - y2    
\State \texttt{inv\_modp} $x_1$ $t_0$;              %// t0 <- 1/(x1 - x2) 
\State \texttt{mul\_modp} $y_1$ $t_0$ $\lambda$;         % // lam <- (y1 - y2)/(x1 - x2)
\State \texttt{mul\_modp} $\lambda$ $x_1$ $y_1$;          %// y1 <- 0
\State \texttt{inv\_modp} $x_1$ $t_0$;              %// t0 <- 0 
\State \texttt{squ\_modp} $\lambda$ $t_0$;             %// t0 <- ((y1 - y2)/(x1 - x2))^2 = lam^2
\State \texttt{ctrl\_sub\_modp} $x_1$ $t_0$ $\ctr$;   % // x1 <- x1 - x2 - ((y1 - y2)/(x1 - x2))^2 // CTRL
\State \texttt{ctrl\_add\_const\_modp} $x_1$ $3x_2$ $\ctr$;    %// x1 <- x1 - x2 - ((y1 - y2)/(x1 - x2))^2 + 3*x2 = -(x3 - x2) // CTRL
\State \texttt{squ\_modp} $\lambda$ $t_0$;             %// t0 <- 0
\State \texttt{mul\_modp} $\lambda$ $x_1$ $y_1$;        %  // y1 <- y3 + y2
\State \texttt{inv\_modp} $x_1$ $t_0$;              %// t0 <- -1/(x3 - x2)
\State \texttt{mul\_modp} $t_0$ $y_1$ $\lambda$;         % // lam <- 0
\State \texttt{inv\_modp} $x_1$ $t_0$;              %// t0 <- 0
\State \texttt{ctrl\_neg\_modp} $x_1$ $\ctr$;      % // x1 <- x3 - x2 // CTRL
\State \texttt{ctrl\_sub\_const\_modp} $y_1$ $y_2$ $\ctr$;  %  // y3 done // CTRL
\State \texttt{add\_const\_modp} $x_1$ $x_2$;      
\end{algorithmic}
\columnbreak
\begin{algorithmic}[0]
\State //\quad $x_1 \gets x_1 - x_2$
\State //\quad $y_1 \gets [y_1 - y_2]_1,[y_1]_0$  
\State //\quad $t_0 \gets 1/(x_1 - x_2) $t
\State //\quad $\lambda \gets [\frac{y_1 - y_2}{x_1 - x_2}]_1, [\frac{y_1}{x_1 - x_2}]_0$
\State //\quad $y_1 \gets 0$
\State //\quad $ t_0 \gets 0 $
\State //\quad $t_0 \gets \lambda^2$
\State //\quad $x_1 \gets [x_1 - x_2 - \lambda^2]_1, [x_1 - x_2]_0$
\State //\quad $x_1 \gets [x_2 - x_3]_1, [x_1 - x_2]_0$ 
\State //\quad $ t_0\gets 0$
\State //\quad $ y_1 \gets [y_3 + y_2]_1, [y_1]_0$
\State //\quad $ t_0 \gets [\frac{1}{x_2 - x_3}]_1, [\frac{1}{x_1 - x_2}]_0$
\State //\quad $\lambda \gets 0$
\State //\quad $t_0 \gets 0$
\State //\quad $x_1 \gets [x_3 - x_2]_1, [x_1-x_2]_0$ 
\State //\quad $y_1 \gets [y_3]_1, [y_1]_0$
\State //\quad $x_1 \gets [x_3]_1, [x_1]_0$ 
\end{algorithmic}
\end{multicols}
\end{algorithm}

\begin{figure}[ht!]
{\scriptsize
\Qcircuit @C=0.6em @R=1.2em { 
\push{\rule{3.5em}{0em}}& & \lstick{\ket{x_1}} & \qw\slash^n  & \gate{-x_2}  & \sgate{\inv}{3} & \qw & \sgate{\mul}{2} & \sgate{\inv}{3} & \qw & \gateout{-} \qwx[3] & \qw & \gate{+3x_2}  & \qw & \sgate{\mul}{2} & \sgate{\inv}{3} & \qw & \sgate{\inv}{3} & \gate{\nega} & \gate{+x_2} & \qw& \rstick{\ket{x_3}}\\
& & \lstick{\ket{\ctr}} & \qw & \ctrl{1} & \qw & \qw & \qw & \qw & \qw & \ctrl{-1}  & \qw & \ctrl{-1}& \qw  & \qw & \qw & \qw & \qw & \ctrl{-1} & \ctrl{1} & \qw & \rstick{\ket{\ctr}}\\
& & \lstick{\ket{y_1}} & \qw\slash^n  & \gate{-y_2} & \qw & \sgate{\mul}{1} & \gateout{\mul} \qwx[2] & \qw & \qw & \qw & \qw & \qw & \qw  & \gateout{\mul} \qwx[2] & \qw & \sgate{\mul}{1} & \qw & \qw & \gate{-y_2} & \qw& \rstick{\ket{y_3}} \\
& & \lstick{\ket{t_0=0}}  & \qw\slash^n  & \qw & \gateout{\inv} & \sgate{\mul}{1} & \qw & \gateout{\inv} & \multigateout{1}{\squ} & \gate{-} & \qw  & \multigateout{1}{\squ} & \qw  & \qw & \gateout{\inv} & \sgate{\mul}{1} & \gateout{\inv} & \qw & \qw & \qw & \rstick{\ket{0}} \\
& & \lstick{\ket{\lambda=0}} & \qw\slash^n & \qw & \qw & \gateout{\ \mul} & \gate{\mul} & \qw & \ghost{\squ} & \qw & \qw & \ghost{\squ} & \qw  & \gate{\mul} & 
\qw & \gateout{\ \mul} & \qw & \qw & \qw & \qw & \rstick{\ket{0}} 
%\gategroup{1}{4}{6}{4}{.7em}{.}
%\gategroup{4}{5}{7}{5}{.7em}{.}
%\gategroup{1}{6}{7}{6}{.7em}{.}
%\gategroup{1}{7}{6}{7}{.7em}{.}
%\gategroup{1}{9}{6}{9}{.7em}{.}
%\gategroup{1}{13}{7}{13}{.7em}{.}
%\gategroup{1}{14}{6}{14}{.7em}{.}
%\gategroup{4}{15}{7}{15}{.7em}{.}
}
}
\caption{Quantum circuit for controlled elliptic curve point addition. All operations are modulo $p$ and we use the abbreviations $+ \gets \mathtt{add\_modp}$, $- \gets \mathtt{sub\_modp}$, $\mul \gets \mathtt{mul\_modp}$, $\squ \gets \mathtt{squ\_modp}$, $\inv \gets \mathtt{inv\_modp}$. }\label{fig:ECpointadd}
\end{figure}
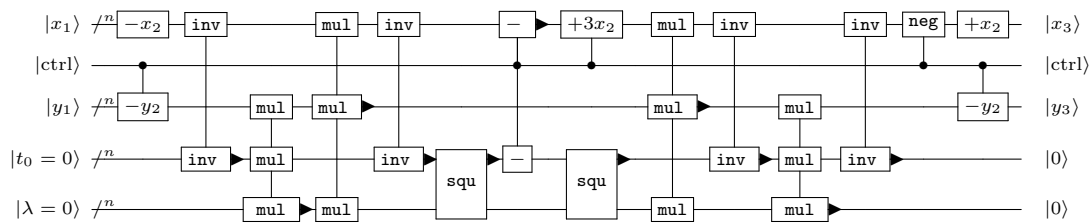

Figure~\ref{fig:ECpointadd} shows a quantum circuit that implements Algorithm~\ref{alg:ECpointadd}. The quantum registers $\ket{x_1}, \ket{y_1}, \ket{t_0}, \ket{\lambda}$ all consist of $n$ logical qubits, whereas $\ket{\ctr}$ is a single logical qubit. For simplicity in the circuit diagram, we do not show the register $\ket{\tmp}$ with the auxiliary qubits. These qubits are used as needed by the modular arithmetic operations and are returned to their original state after each operation. The largest amount of ancilla qubits is needed by the modular inversion algorithm, which determines that we require $5n$ qubits in the register $\ket{\tmp}$. To avoid permuting the wires between gates, we have used a split gate notation for some modular operations. For all gates, the black triangles mark the output wire that contains the result. As described in Section~\ref{sec:mod}, addition and subtraction gates carry out their operations in place, meaning that one of the input registers is overwritten with the result. Modular multiplication, squaring and inversion operate out of place and store the result in a separate output register.

\begin{remark}(Projective coordinates)
As can be seen from Section~\ref{sec:mod}, modular inversion is by far the most complex and resource consuming part of the elliptic curve point addition. The need for computing and uncomputing the slope $\lambda$ leads to four calls to the inversion in Algorithm~\ref{alg:ECpointadd}. In accordance with the observations provided in~\cite{PZ03}, it accounts for the main cost of the algorithm.

Unsurprisingly, this situation resembles the one for classical modular arithmetic. For example, in elliptic curve cryptography, a modular inversion can be two orders of magnitudes more costly than a modular multiplication, depending on the specific prime field. A significant speed-up can be achieved by using some form of projective coordinates\footnote{A collection of possible coordinate systems and the corresponding formulas to carry out the group law is provided at \url{https://www.hyperelliptic.org/EFD/}.}, which allow to avoid almost all modular inversions in cryptographic protocols by essentially multiplying through with all denominators. This comes at the relatively small cost of storing more coefficients and a moderate increase in addition and multiplication operations and has proved highly effective. It is thus a natural question to ask whether the use of projective coordinates can also make Shor's algorithm more efficient. 

There are several obstacles that make it non-trivial to use projective coordinates for quantum algorithms, such as the fact that each point is represented by an equivalence class of coordinate vectors and the increased number of temporary variables, which need to be uncomputed~\cite{MMCP09}. 
In this work, we thus refrained from investigating projective coordinate systems any further and leave it as an open problem to explore their benefits in the context of Shor's algorithm.
\end{remark}

\subsection{Scalar multiplication}
In order to compute a scalar multiplication $[m]P$ of a known base point $P$, we also follow the approach outlined in~\cite{PZ03}. Namely, by classically precomputing all $n$ $2$-power multiples of $P$, the scalar multiple can be computed by a sequence of $n$ controlled additions of those constant points to an accumulator in a quantum register along the binary representation of the scalar. Namely, let $m=\sum_{i=0}^{n-1}m_i2^i$, $m_i \in \{0,1\}$, be the binary representation of the $n$-bit scalar $m$. Then,
\[[m]P = [\sum_{i=0}^{n-1}m_i2^i]P = \sum_{i=0}^{n-1}m_i([2^i]P).\]
This has the advantage that all doubling operations can be carried out on a classical computer and the quantum circuit only requires the generic point addition, which simplifies the overall implementation.

The latter has been argued by Proos and Zalka in~\cite[Section~4.2]{PZ03}. They say that on average, for any addition step, the probability of an exceptional case is negligibly low, and hence this will only have a negligible influence on the fidelity of the algorithm. To prevent the addition with the point at infinity in the first step, they suggest to initialize the register with a non-zero multiple of the point $P$. For the purpose of estimating resources for Shor's algorithm, we follow the approach by Proos and Zalka and only consider the generic group law. We will have a closer look at the details next.

\subsubsection{Counting scalars with exceptional cases.} As explained in Section~\ref{sec:ecshor}, Shor's algorithm involves generating a superposition over all possible pairs of $(n+1)$-bit strings $k$ and $\ell$, i.e. the state $\frac{1}{2^{n+1}} \sum_{k,\ell=0}^{2^{n+1}-1} \ket{k,\ell}$. Then over this superposition, involving two additional $n$-qubit registers to hold an elliptic curve point, one computes a double scalar multiplication $\frac{1}{2^{n+1}} \sum_{k,\ell=0}^{2^{n+1}-1} \ket{k,\ell} \ket{[k]P+[\ell]Q}$ of the input points given by the ECDLP instance. 

Figure~\ref{fig:shorcircuit} depicts the additional elliptic curve point register to be initialized with a representation of the neutral element $\O$. But if we only consider the generic case of the group law, the first addition of $P$ would already involve an exceptional case due to one of the inputs being $\O$. Proos and Zalka~\cite{PZ03} propose to solve this issue by instead initializing the register with a uniform random non-zero multiple of $P$, say $[a]P$ for a random $a\in \{1,2,\dots,r-1\}$. Recall that $r$ is the order of $P$ which we assume to be a large prime. Now, if $a\notin\{1,r-1\}$, the first point addition with $P$ works as a generic point addition. With high probability, this solves the issue of an exception in the first addition, but still exceptions occur along the way for many of the possibilities for bit strings $k$ and $\ell$. Whenever a bit string leads to an exceptional case in the group law, it produces a wrong result for the double scalar multiplication and pollutes the quantum register. We call such a scalar invalid. For Shor's algorithm to work, the overall number of such invalid scalars must be small enough. In the following we count these scalars, similar to the reasoning in~\cite{PZ03}. 

\subsubsection{Exceptional additions of a point to itself.} Let $a \in \{1,2,\dots, r-1\}$ be fixed and write $k = \sum_{i=0}^{n}k_i2^i$, $k_i\in \{0,1\}$. We first consider the exceptional case in which both input points are the same, which we call an exceptional doubling. If $a=1$, this occurs in the first iteration for $k_0=1$, because we attempt to add $P$ to itself. This means that for $a=1$, all scalars $k$ with $k_0=1$ lead to a wrong result and therefore half of the scalars are invalid, i.e. in total $2^n$. 

For $a=2$, the case $k_0=1$ is not a problem since the addition $[2]P + P$ is a generic addition, but $(k_0, k_1) = (0,1)$ leads to an exceptional doubling operation in the second controlled addition. This means that all scalars $(0,1,k_2,\dots,k_n)$ are invalid. These are one quarter of all scalars, i.e. $2^{n-1}$. 

For general $a$, assume that $k$ is a scalar such that the first $i-1$ additions, $i \in \{1,\dots,n\}$, controlled on the bits $k_0,\dots,k_{i-1}$ do not encounter any exceptional doubling cases. The $i$-th addition means the addition of $[2^i]P$ for $0\leq i \leq n$. Then the $i$-th addition is an exceptional doubling if, and only if
\[
a + (k_0 + k_1\cdot 2 + \dots + k_{i-1}\cdot 2^{i-1}) = 2^i\!\pmod r.
\]
If $i$ is such that $2^i < r$. Then, the above condition is equivalent to the condition $a = 2^i - \sum_{j=0}^{i-1}k_j\cdot 2^j$ over the integers. This means that an $a$ can only lead to an exceptional doubling in the $i$-th addition if $a \in \{1,\dots,2^i\}$. Furthermore, if $i$ is the smallest integer, such that there exist $k_0, \dots, k_{i-1}$ such that this equation holds, we can conclude that $a\in \{2^{i-1}+1,\dots,2^i\}$ and $k_{i-1}=0$. In that case, any scalar of the form $(k_0,\dots,k_{i-2},0,1,*,\dots,*)$ is invalid. The number of such scalars is $2^{n-i}$.

If $i$ is instead such that $2^i \ge r$ and if $a\le 2^i-\mu r$ for some positive integer $\mu \leq \lfloor 2^i/r \rfloor$, then in addition to the solutions given by the equation over the integers as above, there exist additional solutions given by the condition $a = (2^i-\mu r) - \sum_{j=0}^{i-1}k_j\cdot 2^j$, namely $(k_0,\dots,k_{i-1},1,*,\dots,*)$. The maximal number of such scalars is $\lfloor (2^i-a)/r\rfloor 2^{n-i}$, but we might have counted some of these already.

For a given $a \in \{1,2,\dots,r-1\}$, denote by $S_a$ the set of scalars that contain an exceptional doubling, i.e. the set of all $k = (k_0, k_1,\dots,k_n)\in\{0,1,\}^{n+1}$ such that there occurs an exceptional doubling when executing the addition $[a + \sum_{j=0}^{i-1}k_j\cdot 2^j]P + [2^i]P$ for any $i \in \{0,1,\dots,n\}$. Let $i_a=\lceil\log(a)\rceil$. Then, an upper bound for the number of invalid scalars is given by 
\[
\#S_a \leq 2^{n-i_a} + \sum_{i=\lceil\log(r)\rceil}^n \lfloor (2^i-a)/r\rfloor 2^{n-i}. 
\]
Hasse's bound gives $\lceil \log(r)\rceil \ge n-1$, which means that
\[
\#S_a \leq 2^{n-i_a} + 2\lfloor (2^{n-1}-a)/r\rfloor + \lfloor (2^n-a)/r\rfloor \leq 2^{n-i_a} + 8. 
\]
Hence on average, the number of invalid scalars over a uniform choice of $k \in \{1,\dots,r-1\}$ can be bounded as
\[
\sum_{a=1}^{r-1}\mathrm{Pr}(a)\cdot \#S_a \leq \frac{1}{r-1}\sum_{a=1}^{r-1} 2^{n-\lceil \log(a)\rceil} + 8.
\]
Grouping values of $a$ with the same $\lceil \log(a) \rceil$ and possibly adding terms at the end of the sum, the first term can be simplified and further bounded by $\frac{1}{r-1}(2^n + \lceil\log(r-1)\rceil 2^{n-1}) = (2+\lceil\log(r-1)\rceil)\frac{2^{n-1}}{r-1}$. For large enough bitsizes, we use that $r-1\geq 2^{n-1}$ and obtain the upper bound on the expected number of invalid scalars of roughly $\lceil\log(r)\rceil + 10 \approx n+10$. This corresponds to a negligible fraction of about $n/2^{n+1}$ of all scalars. 

\subsubsection{Exceptional additions of a point to its negative.} To determine the number of invalid scalars arising from the second possibility of exceptions, namely the addition of a point to its negative, we carry out the same arguments. An invalid scalar is a scalar that leads to an addition $[-2^i]P + [2^i]P$. The condition on the scalar $a$ is slightly changed with $2^i$ replaced by $r-2^i$, i.e.
\[
a + (k_0 + k_1\cdot 2 + \dots + k_{i-1}\cdot 2^{i-1}) = r-2^i\!\pmod r.
\]
Whenever this equation holds over the integers, i.e. $r-a = 2^i + (k_0 + k_1\cdot 2 + \dots + k_{i-1}\cdot 2^{i-1})$ holds, we argue analogously as above. If $2^i < r$ and $r-a \in \{2^i,\dots,2^{i+1}-1\}$, there are $2^{n-i}$ invalid scalars. Similar arguments as above for the steps such that $2^i >r$ lead to similar counts. Overall, we conclude that in this case the fraction of invalid scalars can also be approximated by $n/2^{n+1}$.

\subsubsection{Exceptional additions of the point at infinity.} Since the quantum register holding the elliptic curve point is initialized with a non-zero point and the multiples of $P$ added during the scalar multiplication are also non-zero, the point at infinity can only occur as the result of an exceptional addition of a point to its negative. Therefore, all scalars for which this occurs have been excluded previously and we do not further consider this case.
\\

\noindent Overall, an approximate upper bound for the fraction of invalid scalars among the superposition of all scalars due to exceptional cases in the addition law is $2n/2^{n+1} = n/2^n$.

\subsubsection{Double scalar multiplication.} In Shor's algorithm with the above modification, one needs to compute a double scalar multiplication $[a+k]P+[\ell]Q$ where $P$ and $Q$ are the points given by the ECDLP instance we are trying to solve and $a$ is a fixed uniformly random non-zero integer modulo $r$. We are trying to find the integer $m$ modulo $r$ such that $Q=[m]P$. Since $r$ is a large prime, we can assume that $m\in \{1,\dots,r-1\}$ and we can write $P = [m^{-1}]Q$. Multiplication by $m^{-1}$ on the elements modulo $r$ is a bijection, simply permuting these scalars. Hence, after having dealt with the scalar multiplication to compute $[a+k]P$ above, we can now apply the same treatment to the second part, the addition of $[\ell]Q$ to this result. 

Let a be chosen uniformly at random. For any $k$, we write $[a+k]P = [m^{-1}(a+k)]Q$. Assume that $k$ is a valid scalar for this fixed choice of $a$. Then, the computation of $[a+k]P$ did not involve any exceptional cases and thus $[a+k]P\neq \O$, which means that $a+k \neq 0\!\pmod r$. If we assume that the unknown discrete logarithm $m$ has been chosen from $\{1,\dots, r-1\}$ uniformly at random, then the value $b = m^{-1}(a+k)\!\mod r$ is uniform random in $\{1,\dots,r-1\}$ as well, and we have the same situation as above when we were looking at the choice of $a$ and the computation of $[a+k]P$. 

Using the rough upper bound for the fraction of invalid scalars from above, for a fixed random choice of $a$, the probability that a random scalar $k$ is valid, is at least $1-n/2^n$. Further, the probability that $(k,\ell)$ is a pair of valid scalars for computing $[a+k]P+[\ell]Q$, conditioned on $k$ being valid for computing $[a+k]P$ is also at least $1-n/2^n$. Hence, for a fixed uniform random $a$, the probability for $(k,\ell)$ being valid is at least $(1-n/2^n)^2 = 1-n/2^{n-1} + n^2/2^{2n}\approx 1-n/2^{n-1}$. This result confirms the rough estimate by Proos and Zalka~\cite[Section~4.2]{PZ03} of a fidelity loss of $4n/p\geq 4n/2^{n+1}$. 

\begin{remark} (Complete addition formulas)
There exist complete formulas for the group law on an elliptic curve in Weierstrass form~\cite{BL95}. This means that there is a single formula that can evaluate the group law on any pair of $\F_p$-rational points on the curve and thus avoids the occurrence of exceptional cases altogether. For classical computations, this comes at the cost of a relatively small slowdown~\cite{RCB16}. Using such formulas would increase the algorithm's fidelity in comparison to the above method. Furthermore, there exist alternative curve models for elliptic curves which allow coordinate systems that offer even more efficient complete formulas. 
One such example is the twisted Edwards form of an elliptic curve~\cite{BBJLP08}. However, not all elliptic curves allow a curve model in twisted Edwards form, like, for example, the prime order NIST curves. We leave it as an open problem to investigate the use of a complete group law, or more generally the use of different curve models and coordinate systems in Shor's ECDLP algorithm. 
\end{remark}

\section{Cost and resource estimates for Shor's algorithm}
\label{sec:shor}
We implemented the reversible algorithm for elliptic curve point addition on elliptic curves $E$ in short Weierstrass form defined over a prime field $\F_p$, where $p$ has $n$ bits, as shown in Algorithm~\ref{alg:ECpointadd} and Figure~\ref{fig:ECpointadd} in Section~\ref{sec:ecadd} in F\# within the quantum computing software tool suite \liq\ ~\cite{liquid}. This allows us to test and simulate the circuit and all its components and obtain precise counts of the number of qubits, the number of Toffoli gates and the Toffoli gate depth for a working simulation. We thus do not have to rely on mere estimates obtained by pen-and-paper calculations and thus gain a higher confidence in the results.
When implementing the algorithms, our overall emphasis was to minimize first the number of required logical qubits and second the Toffoli gate count. 

We have simulated and tested our implementation for cryptographically relevant parameter sizes and were able to simulate the elliptic curve point addition circuit for curves over prime fields of size up to $521$ bits.  
For each case, we computed the number of qubits required to implement the circuit, and its size and depth in terms of Toffoli gates.

\subsubsection{Number of logical qubits.}
The number of logical qubits of the modular arithmetic circuits in our simulation that are needed in the elliptic curve point addition are given in Table~\ref{tbl:modqubits}. We list each function with its total required number of qubits and the number of ancilla qubits included in that number. 
All ancilla qubits are expected to be input in the state $\ket{0}$ and are returned in that state, except for the circuits in the first two rows, which only require one or two such ancilla qubits and $n-1$ or $n-2$ ancillas in an unknown state to which they will be returned.
The addition, subtraction and negation circuits all work in place, such that one $n$-qubit input register is replaced with the result. The multiplication, squaring and inversion circuits require an $n$-qubit register with which the result of the computation is XOR-ed. 

\begin{table}[ht!]
\centering
{\footnotesize
\renewcommand{\tabcolsep}{0.3cm}
\renewcommand{\arraystretch}{1.1}
\begin{tabular}{|l@{\!}|c|c|c|}
\hline
Modular arithmetic circuit & \multicolumn{2}{c|}{$\#$ of qubits} & $\#$ Toffoli\\
& total & ancillas & gates\\
\hline\hline
\begin{minipage}{3cm} \smallskip \texttt{add\_const\_modp}, \texttt{sub\_const\_modp} \smallskip \end{minipage} & $2n$ & $n$ & $16n\log_2(n)-26.9 n$ \\ \hline
\begin{minipage}{3cm} \smallskip \texttt{ctrl\_add\_const\_modp}, \texttt{ctrl\_sub\_const\_modp} \smallskip \end{minipage} & $2n+1$ & $n$  &$16n\log_2(n)-26.9 n$  \\  \hline
\begin{minipage}{4cm} \medskip \texttt{ctrl\_sub\_modp} \medskip \end{minipage} & $2n+4$ & $3$ & $16 n \log_2(n) - 23.8 n$\\ \hline
\begin{minipage}{4cm} \medskip \texttt{ctrl\_neg\_modp} \medskip \end{minipage}& $n+3$ & $2$ & $8n \log_2(n) - 14.5 n$\\ \hline
\begin{minipage}{4cm} \medskip \texttt{mul\_modp} (dbl/add) \medskip \end{minipage}& $3n+2$ & $2$ & $32 n^2 \log_2(n) -59.4 n^2$\\ \hline
\begin{minipage}{4cm} \medskip \texttt{mul\_modp} (Montgomery) \medskip \end{minipage}& $5n+4$ & $2n+4$ & $16 n^2 \log_2(n) - 26.3 n^2$ \\ \hline
\begin{minipage}{4cm} \medskip \texttt{squ\_modp} (dbl/add) \medskip \end{minipage}& $2n+3$ & $3$ & $32 n^2 \log_2(n) -59.4 n^2$\\ \hline
\begin{minipage}{4cm} \medskip \texttt{squ\_modp} (Montgomery) \medskip \end{minipage}& $4n+5$ & $2n+5$ & $16 n^2 \log_2(n) - 26.3 n^2$\\ \hline
\begin{minipage}{4cm} \medskip \texttt{inv\_modp} \medskip \end{minipage} & $7n+2\lceil\log_2(n)\rceil+9$ & $5n+2\lceil\log_2(n)\rceil+9$ & $32 n^2 \log_2(n)$\\ 
\hline
\end{tabular}
\vspace{1em}
}
\caption{Total number of qubits and number of Toffoli gates needed for the modular arithmetic circuits used in the elliptic curve point addition on $E/\F_p$ with $n$-bit prime $p$. The column labeled ``ancilla'' denotes the number of required ancilla qubits included in the total count. Except for the first two rows (un-/controlled constant addition/subtraction), they are expected to be input in state $\ket{0\dots 0}$ and are returned in that state. The constant addition/subtraction circuits in the first row only need one clean ancilla qubit and can take $n-1$ dirty ancilla qubits in an unknown state, in which they are returned. The controlled constant addition/subtraction circuits in the second row use two dirty ancillas. \label{tbl:modqubits}}
\end{table}

Although the modular multiplication circuit based on modular doubling and additions uses less qubits than Montgomery multiplication, we have used the Montgomery approach to report the results of our experiments. Since the lower bound on the overall required number of qubits is dictated by the modular inversion circuit, neither multiplication approach adds qubit registers to the elliptic curve addition circuit since they can use ancilla qubits provided by the inversion algorithm. We therefore find that the Montgomery circuit is the better choice then because it reduces the number of Toffoli gates substantially.    

Because the maximum amount of qubits is used during an inversion operation, the overall number of logical qubits for the controlled elliptic curve point addition in our simulation is 
\[9n+2\lceil\log_2(n)\rceil+10.\] 
In addition to the $7n+2\lceil\log_2(n)\rceil+9$ required by the inversion, an additional qubit is needed for the control qubit $\ket{\ctr}$ of the overall operation and $2n$ more qubits are needed since two $n$-qubit registers need to hold intermediate results during each inversion. 

\subsubsection{Number of Toffoli gates and depth.}
Perhaps surprisingly, the precise resource count of the number of Toffoli gates in the constructed circuits is not a trivial matter. There are two main reasons for this: first, as constants are folded, the actual value of the constants matter as different bit-patterns (e.g., of the underlying prime $p$) give rise to different circuits. This effect, however, is not large and does not impact the leading order coefficients for the functions in the table. Second, the asymptotically dominating cost arises from the incrementer construction based on \cite{HRS16}. For the basic functions reported in Table \ref{tbl:modqubits} one can determine the number of incrementers used, which is either of the form $a n \log_2(n)$ or $a n^2 \log_2(n)$ with a constant $a$. We determined the leading order term by inspection of the circuit and determining how many constant incrementers occur. Then we computed a regression of the next order term by solving a standard polynomial interpolation problem. The results are summarized in the last column of Table \ref{tbl:modqubits}.  

\begin{figure}[t!]
\begin{tabular}{c@{\;}c}
\!\!\!\!
\raisebox{0.1mm}{\includegraphics[height=4.2cm]{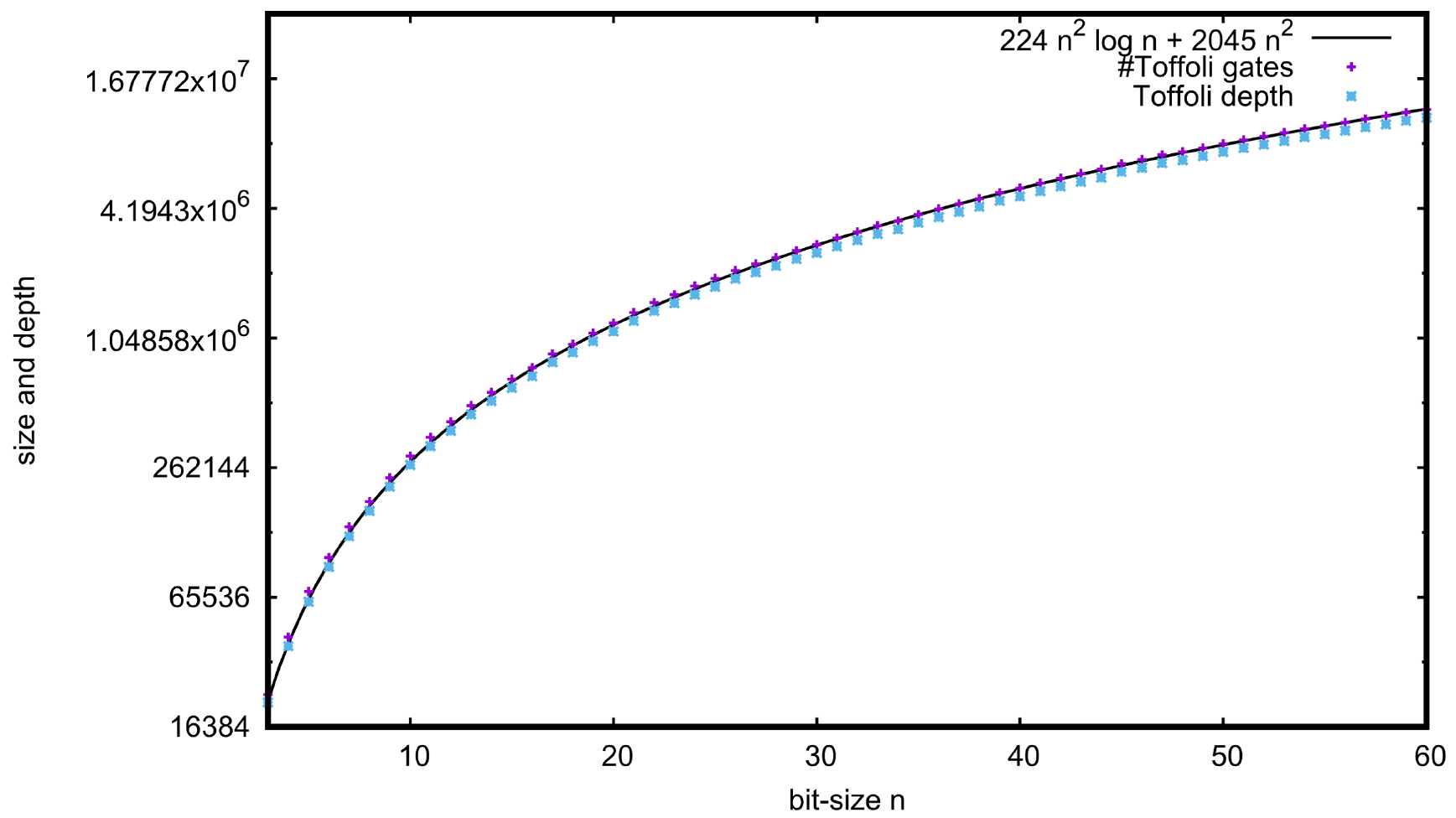}} &
\includegraphics[height=4.2cm]{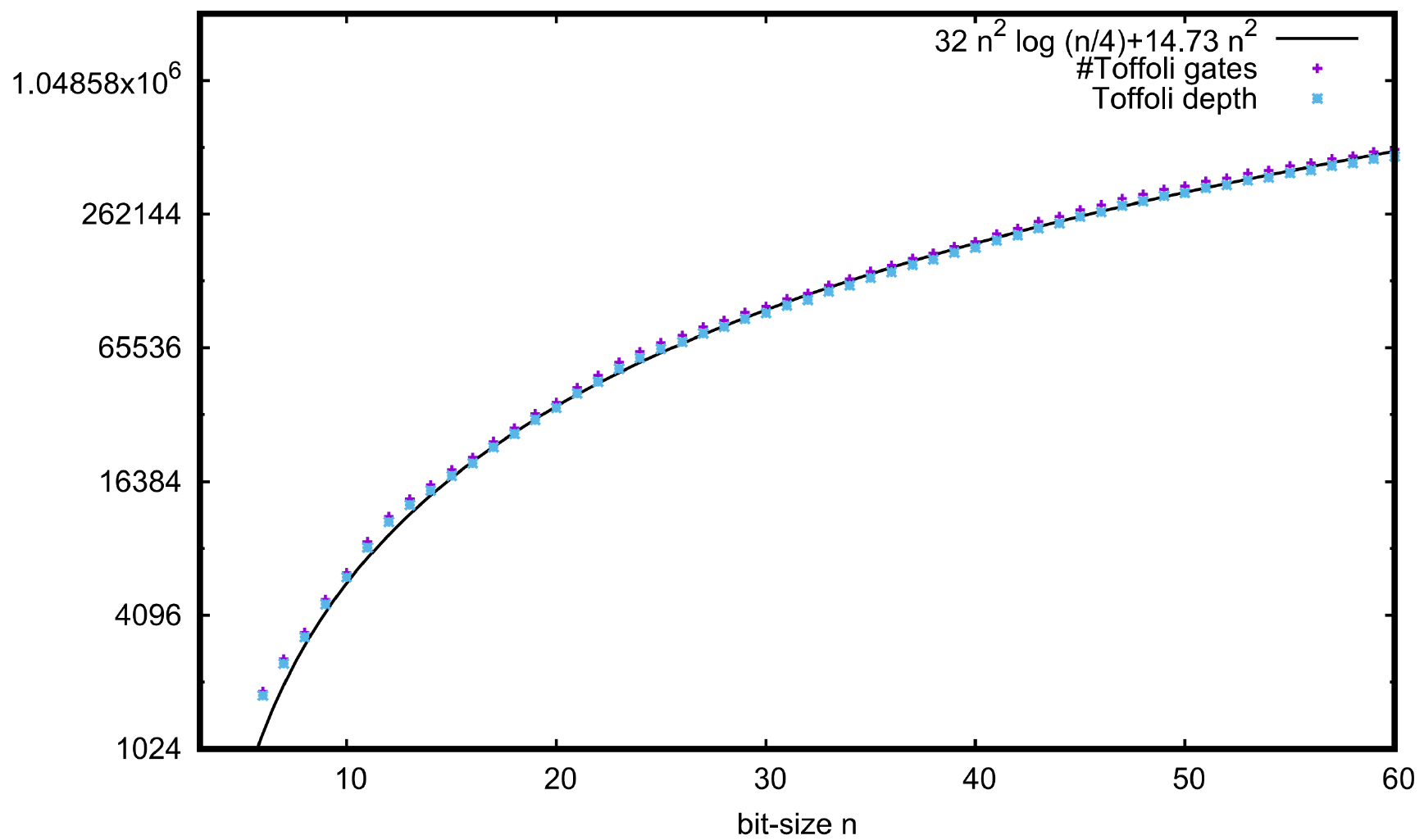} 
\end{tabular}
\caption{Shown on the left are resource estimates for the number of Toffoli gates and the Toffoli gate depth for the implementation of elliptic curve point addition $\ket{P} \mapsto \ket{P+Q}$, where $Q$ is a constant point. Shown on the right are resource estimates for the same metrics for modular multiplication $\ket{x} \mapsto \ket{ax \; {\rm mod} \; N}$, where $a$ and $N$ are constants. Fitting the data for the elliptic curve case we obtain a scaling as $224 n^2 \log_2(n)+2045 n^2$ up to lower order terms. The cost for the entire Shor algorithm over the elliptic curve scales as $2n$ the cost for a single point addition, i.e. $448 n^3 \log_2(n)$ up to lower order terms. As shown in \cite{HRS16}, the cost for modular multiplication scales as $32n^2 (\log_2(n)-2)+14.73 n^2$ and the cost of the entire Shor factoring algorithm scales as $64n^2 \log_2(n)$. \label{fig:estimates}}
\end{figure}

Putting everything together, we now obtain an estimate for the entire group law as computed by Algorithm \ref{alg:ECpointadd}: As there are a total of $4$ inverters, $2$ squarers, and $4$ multipliers, we obtain that the leading order coefficient of a single point addition is $224 = 4 \cdot 32 + 2 \cdot 16 + 4 \cdot 16$. We then again perform a regression to determine the next coefficient. As a result, we estimate that the number of Toffoli gates in the point addition circuit scales as $224 n^2 \log_2(n)+ 2045 n^2$. Figure~\ref{fig:estimates} shows the scaling of the estimates for the Toffoli gate count and the Toffoli gate depth of the circuit for a range of relatively small bit sizes $n$. To estimate the overall resource requirements for Shor's algorithm, one simply multiplies by $2n$, since the controlled point addition is iterated $2n$ times. This leads to the overall estimate for the scaling of the number of Toffoli gates in Shor's ECDLP algorithm as
\[(448\log_2(n)+4090) n^3.\]

With respect to a given circuit, the Toffoli depth is computed as follows: we sweep all gates in the circuits and keep a running counter for each qubit on which time step it was acted upon last by a Toffoli gate. The depth is then the maximum of these quantities over all qubits. As the number of qubits is comparatively small in the circuits considered here, we can perform these updates efficiently, leading to an algorithm to compute the depth in time linear in the number of gates in the circuit. Note that whenever we encounter a CNOT or NOT gate, we do not increase the counter as by our assumption these gates do not contribute to the overall depth as they are Clifford gates. Overall, we find that the circuit Toffoli depth is a little bit smaller than the total number of Toffoli gates which shows that there is some parallelism in the circuit that can be exploited when implementing it on a quantum computer that facilitates parallel application of quantum gates. 

We compare our results to the corresponding simulation results for Shor's factoring algorithm presented in~\cite{HRS16}, where the corresponding function is modular constant multiplication. In this case, the number of Toffoli gates scales as $32n^2 (\log_2(n)-2)+14.73 n^2$, where $n$ is the bitsize of the modulus to be factored. As above, to estimate the overall resource requirements, one again multiplies by $2n$, which gives $(64 (\log_2(n)-2)+29.46) n^3$.

\begin{table}[hbt]
\begin{center}
\renewcommand{\tabcolsep}{0.3cm}
\renewcommand{\arraystretch}{1.2}
\begin{tabular}{|c|c|c|c|c||c|c|c|}
\hline
\multicolumn{5}{|c||}{ECDLP in $E(\F_p)$} & \multicolumn{3}{c|}{Factoring of RSA modulus $N$} \\
\multicolumn{5}{|c||}{simulation results} & \multicolumn{3}{c|}{interpolation from~\cite{HRS16}} \\
\hline
\!\!\!$\lceil\log_2(p)\rceil$ \!\!\!\! & \!\!\#Qubits\!\! & \#Toffoli & Toffoli & Sim time & \!\!\!$\lceil\log_2(N)\rceil$ \!\!\!\!& \!\!\#Qubits\!\! & \#Toffoli \\
bits & & gates & depth & sec & bits &  & gates\\
\hline\hline
110  & 1014 & $9.44 \cdot 10^{9}$  & $8.66 \cdot 10^{9}$ & 273  &512  & 1026  & $6.41 \cdot 10^{10}$   \\
160  & 1466 & $2.97 \cdot 10^{10}$ & $2.73 \cdot 10^{10}$ & 711  & 1024 & 2050  & $5.81 \cdot 10^{11}$   \\
192 & 1754 & $5.30\cdot 10^{10}$ & $4.86 \cdot 10^{10}$ & 1\,149 & $-$ & $-$ & $-$ \\
224  & 2042 & $8.43 \cdot 10^{10}$ & $7.73 \cdot 10^{10}$ & 1\,881 & 2048 & 4098  & $5.20 \cdot 10^{12}$  \\
256  & 2330 & $1.26 \cdot 10^{11}$ & $1.16 \cdot 10^{11}$ & 3\,848 & 3072 & 6146  & $1.86 \cdot 10^{13}$  \\
384  & 3484 & $4.52 \cdot 10^{11}$ & $4.15 \cdot 10^{11}$ & 17\,003 & 7680 & 15362 & $3.30 \cdot 10^{14}$  \\
521  & 4719 & $1.14 \cdot 10^{12}$ & $1.05 \cdot 10^{12}$ & 42\,888 & 15360 & 30722 & $2.87 \cdot 10^{15}$\\
\hline
\end{tabular}
\bigskip
\caption{Resource estimates of Shor's algorithm for computing elliptic curve discrete logarithms in $E(\F_p)$ versus Shor's algorithm for factoring an RSA modulus $N$.\label{tbl:estimates}}
\end{center}
\end{table}

Table~\ref{tbl:estimates} contains the resources required in our simulated circuits for parameters of cryptographic magnitude that are used in practice. The simulation time only refers to our implementation of the elliptic curve group law. The simulation timings were measured when running our \liq~implementation on an HP ProLiant DL580 Gen8 machine consisting of 4 Intel Xeon processors @ 2.20 GHz and 3TB of memory. The rows for ECC and RSA are aligned so that the corresponding parameters provide a similar classical security level according to NIST recommendations from 2016. 
%A resource estimate for the number of Toffoli gates in the entire Shor algorithm for solving the ECDLP is obtained by regression as $448 n^3 \log_2(n) + 4090 n^3$. Resource estimates for factoring are according to the interpolation $(64 (\log_2(n)-2)+29.46) n^3$~\cite{HRS16}. Simulation time only refers to simulation of a single elliptic curve point addition.

\begin{remark}
Recently, a variation of Shor's quantum algorithms for computing discrete logarithms and factoring was developed in \cite{EH:2017}. The basic observation of this paper is that the quantum circuit size can be reduced at the cost of a more expensive classical post-processing. Inasmuch as the argument given in \cite{EH:2017} applies only to the case of discrete logarithms that are guaranteed to be small, it does not apply to the ECC column in Table \ref{tbl:estimates}, however, the argument given about factoring {\em does} apply. This means that if one is willing to perform a classical post-processing based on lattice enumeration algorithms, one can reduce the rounds of phase estimation from $2n$ to $n/2$. While this does not save qubits, it does lead to a division by $4$ of all reported resource counts on the number of Toffoli gates in the RSA column of Table \ref{tbl:estimates}.
\end{remark}

\begin{remark}
In \cite{BBM:2017} it was shown that even in the presence of a quantum computer with limited size, the heuristic time complexity of the Number Field Sieve can be reduced from $L^{1.901 + o(1)}$ to $L^{1.386 + o(1)}$. Specifically, it is shown that a sub-linear number of $n^{2/3+o(1)}$ qubits is enough for a hybrid quantum-classical algorithm to work. Taking this result into account, shifts the alignment of security parameters in Table \ref{tbl:estimates}, however, it does so in a way that significantly complicates the analysis as constants for the algorithm given in \cite{BBM:2017} would have to be worked out.  
\end{remark}

\section{Discussion}

Comparing to the theoretical estimates by Proos and Zalka in~\cite{PZ03}, our results provide additional evidence that for cryptographically relevant sizes, elliptic curve discrete logarithms can be computed more easily on a quantum computer than factoring an RSA modulus of similar classical security level. However, neither the Toffoli gate counts for factoring that were provided in \cite{HRS16}, nor the ones for elliptic curves that were provided here are as low as the theoretically predicted ``time'' estimates in~\cite{PZ03}.  Also, the number of qubits in our simulation-based estimates is higher than the ones conjectured in~\cite{PZ03}. 

The reasons for the larger number of qubits lie in the implementation of the modular inversion algorithm. Proos and Zalka describe a version of the standard Euclidean algorithm which requires divisions with remainder. We chose to implement the binary GCD algorithm, which only requires additions, subtractions and binary bit shifts. One optimization that applies to both algorithms is register sharing as proposed in~\cite[Section~5.3.5]{PZ03}. The standard Euclidean algorithm as well as the binary GCD work on four intermediate variables, requiring $4n$ bits in total. In our description in Section~\ref{sec:modinv}, these are the variables $u,v,r,s$.  However, Proos and Zalka use a heuristic argument to show that they actually only need about $2n+8\sqrt{n}$ bits at any time during the algorithm. A complication for implementing this optimization is that the boundaries between variables change during the course of the algorithm. We leave it for future work to implement and simulate a reversible modular inversion algorithm that makes use of register sharing to reduce the number of qubits.

Since the basis for register sharing in~\cite{PZ03} is an experimental analysis, Proos and Zalka provide a space analysis that does not take into account the register sharing optimization. With this space analysis, we still need about $2n$ qubits more than their Euclidean algorithm. These qubits come from the fact that our extended binary GCD algorithm generates one bit of garbage in each of the $2n$ rounds. In contrast, \cite{PZ03} only needs $n$ carry qubits. Furthermore, we need an additional $n$-qubit register to copy out the result and run the algorithm in reverse to clean-up all garbage and ancilla qubits. We could not see how to avoid this and how to achieve step-wise reversibility for the extended binary Euclidean algorithm. We leave it as a future challenge to match or even lower the number of qubits for reversible modular inversion from~\cite{PZ03}. 

To summarize, we presented quantum circuits to implement Shor's algorithm to solve the ECDLP. We analyzed the resources required to implement these circuits and simulated large parts of them on a classical machine. Indeed, the overwhelming majority of gates in our circuits are Toffoli, CNOT, and NOT gates, which implement the controlled addition of an elliptic curve point that is known at circuit generation time. We were able to classically simulate the point addition circuit and hence test it for implementation bugs. Our findings imply that attacking elliptic curve cryptography is indeed easier than attacking RSA, even for relatively small key sizes.  

\subsubsection{Acknowledgments.}
We thank Christof Zalka for feedback and discussions and the anonymous reviewers for their valuable comments.

\end{document}